\documentclass[11pt]{article}   	% use "amsart" instead of "article" for AMSLaTeX format
\usepackage{geometry}                		% See geometry.pdf to learn the layout options. There are lots.
\usepackage{graphicx}				% Use pdf, png, jpg, or eps§ with pdflatex; use eps in DVI mode
\usepackage{amssymb}
\usepackage{amsmath}
\usepackage{authblk}
\usepackage[table]{xcolor}
\usepackage{multirow}
\usepackage{graphicx}
\usepackage[normalem]{ulem}

\newcommand{\be}{\begin{equation}}
\newcommand{\ee}{\end{equation}}
\newcommand{\pom}{{I\!\!P}}
\newcommand{\regg}{{I\!\!R}}

\usepackage[utf8]{inputenc}
\usepackage{xspace}
\usepackage[hypertexnames,%a4paper,%
    pdftex,%
    colorlinks,%
		citecolor=blue,%
    hyperindex,%
    plainpages=false,%
    bookmarksopen,%
    bookmarksnumbered%
  ]{hyperref}
\usepackage[numbers,sort&compress]{natbib}
\newcommand\Eq[1]{(\ref{#1})}
\newcommand\Fig[1]{Fig.~\ref{#1}}

\DeclareRobustCommand\GeV{\ensuremath{\mathrm{GeV}}\xspace}

\DeclareRobustCommand\esys{\ensuremath{\delta_\mathrm{sys}}\xspace}
\DeclareRobustCommand\sred{\ensuremath{\sigma_\mathrm{red}}\xspace}
\DeclareRobustCommand\srD{\ensuremath{\sigma_\mathrm{red}^\mathrm{D(3)}}\xspace}

\DeclareRobustCommand\FLDALL{\ensuremath{F_\mathrm{L}^\mathrm{D}}\xspace}
\DeclareRobustCommand\FDALL{\ensuremath{F_\mathrm{2}^\mathrm{D}}\xspace}
\DeclareRobustCommand\FLD{\ensuremath{F_\mathrm{L}^\mathrm{D(3)}}\xspace}
\DeclareRobustCommand\FtwoD{\ensuremath{F_2^\mathrm{D(3)}}\xspace}
\DeclareRobustCommand\FTD{\ensuremath{F_\mathrm{T}^\mathrm{D(3)}}\xspace}
\DeclareRobustCommand\YL{\ensuremath{Y_\mathrm{L}}\xspace}

\DeclareRobustCommand\CM{\ensuremath{\mathrm{CM}}\xspace}
\DeclareRobustCommand\EsetAll{S-17\xspace}
\DeclareRobustCommand\EsetBig{S-9\xspace}
\DeclareRobustCommand\EsetMin{S-5\xspace}
\DeclareRobustCommand\xL{\ensuremath{x_\mathrm{L}}\xspace}

\newcommand\DD{\mathrm{D}}

\definecolor{eM}{hsb}{0.3,0.5,1}
\definecolor{eS}{hsb}{0,1,0.6}

\newcommand\sBD[1]{\cellcolor{eM}\color{eS}\textbf{#1}}
\newcommand\sB[1]{\color{eS}\textbf{#1}}

\graphicspath{{figs/}}
\usepackage{color}
\usepackage[normalem]{ulem}
\definecolor{pncolor}{rgb}{0,0.1,0.7}
\definecolor{ascolor}{rgb}{1,0,1}
\definecolor{nacolor}{rgb}{1,0,0}
\definecolor{wscolor}{rgb}{0,0.6,0.2}

\DeclareRobustCommand\pnout{\bgroup\markoverwith{\color{pncolor}{\rule[0.4ex]{2pt}{0.8pt}}}\ULon}
\DeclareRobustCommand\asout{\bgroup\markoverwith{\color{ascolor}{\rule[0.4ex]{2pt}{0.8pt}}}\ULon}
\DeclareRobustCommand\naout{\bgroup\markoverwith{\color{nacolor}{\rule[0.4ex]{2pt}{0.8pt}}}\ULon}
\DeclareRobustCommand\wsout{\bgroup\markoverwith{\color{wscolor}{\rule[0.4ex]{2pt}{0.8pt}}}\ULon}

\title{\bf Diffractive longitudinal structure function at the Electron Ion Collider}
\author[1]{N\'estor Armesto}
\author[2]{Paul R. Newman}
\author[3]{Wojciech S\l{}omi\'nski}
\author[4]{ Anna M. Sta\'sto}
\affil[1]{\small \it Instituto Galego de F\'{\i}sica de Altas Enerx\'{\i}as IGFAE,
Universidade de Santiago de Compostela, 15782 Santiago de Compostela, Galicia-Spain}
\affil[2]{\small \it School of Physics and Astronomy, University of Birmingham, UK}
\affil[3]{\small \it Institute of Theoretical Physics, Jagiellonian University, Kraków, Poland}
\affil[4]{\small \it Department of Physics, Penn State University, University Park, PA 16802, USA}

\begin{document}
\maketitle
\begin{abstract}
%\pncomment{New suggested abstract compiled from all suggestions.}
Possibilities for the measurement of the longitudinal  
structure function in diffraction \FLDALL\ at the future US
Electron Ion Collider are investigated. The sensitivity to \FLDALL\ arises from the 
variation of the reduced diffractive cross section with centre-of-mass
energy. 
Simulations are performed with various sets of 
beam energy combinations and for different assumptions on the precision
of the diffractive cross section measurements. 
Scenarios compatible with current EIC performance expectations lead to 
an unprecedented 
precision on \FLDALL at the 5-10\% level in the best measured
regions. 
While scenarios with data at 
a larger number of centre-of-mass energies allow the extraction 
of \FLDALL in the widest kinematic domain and with the smallest 
uncertainties, even the more conservative assumptions
lead to precise measurements.  
The ratio $R^\DD$ of photoabsorption cross sections for longitudinally to
transversely polarised photons can also be obtained with high precision
using a separate extraction method.

\end{abstract}

% =================================================
\section{Introduction}
\label{sec:intro}

%\ascomment{ Check the titles of the sections/subsections.   }\nacomment{I find them fine.} \pncomment{Me too}

Diffraction in deep inelastic scattering (DIS)  was studied extensively at the Hadron-Elektron-Ring\-an\-lage (HERA) collider at DESY. The    measurements showed that it gives a large contribution, of about $\sim 10$\%,  to the total cross section ~\cite{Adloff:1997sc,Breitweg:1997aa}, see the review~\cite{Newman:2013ada} and refs. therein. Diffractive events are characterised by the measurement of either a proton (in the case of coherent diffraction) or a state with the proton quantum numbers (incoherent diffraction). Experimentally, diffractive events are defined either by the identification of a proton in dedicated 
far-forward detectors 
housed in Roman pot insertions to the beampipe 
(see for example \cite{Aktas:2006hx, Aaron:2010aa,ZEUS:2004luu,ZEUS:2008xhs}), or by a lack of hadronic activity in 
a sizeable kinematic region 
adjacent to the outgoing proton beam,
i.e. the presence of a large rapidity gap (LRG) (see for example \cite{ZEUS:2008xhs,H1:2006zyl,H1:2012pbl}). Based on the experimental results from HERA, it was possible to  analyse the 
partonic structure of the $t$-channel colourless exchange in such events. A successful description of the diffractive structure functions was achieved at high $Q^2$ based on the collinear factorization and Dokshitzer-Gribov-Lipatov-Altarelli-Parisi (DGLAP) evolution of the   corresponding diffractive parton densities (DPDF) \cite{Aktas:2006hx,Chekanov:2009aa}. The latter quantities parametrize the partonic content of the colourless exchange in the diffractive events.

Diffraction has been a central subject in investigations of strong interactions 
for many decades~\cite{Kaidalov:1979jz,Kaidalov:2003vg}. As a pure quantum phenomenon, some properties derive from basic requirements like unitarity. On the other hand, the microscopic dynamics by which a composite object, a hadron or nucleus, is able to undergo a high energy collision and remain colourless with its constituents bound, is closely related to the confinement mechanism  \cite{Bartels:2000ze}. Besides, diffraction is very sensitive to the high energy behaviour of Quantum Chromodynamics (QCD), specifically to the low-$x$ distribution of partons and its energy evolution  \cite{Kovchegov:2012mbw}. Therefore, it is a promising observable for observing deviations from linear evolution like higher twist effects or parton saturation. 
Diffractive $ep$ scattering is also related to
nuclear shadowing on deuterons~\cite{Gribov:1968jf}
and tests the validity of perturbative factorisation~\cite{Collins:1997sr,Berera:1995fj,Trentadue:1993ka} -- known to be violated in diffractive dijet photoproduction~\cite{Klasen:2008ah}. 
Furthermore, due to the simplicity of the final state, 
diffractive events may offer new opportunities for the detection 
of rare phenomena,
see~\cite{LHCForwardPhysicsWorkingGroup:2016ote} and refs. therein.

Among the various diffractive observables that can be measured in DIS,  a very interesting one, yet experimentally challenging, is  the diffractive longitudinal  structure function \FLDALL. Given by the coupling of  virtual photons with longitudinal polarisation to the hadron that undergoes the diffractive interaction, it is -- as in the case of inclusive events -- a more sensitive probe of the gluon content 
of the target and of the QCD evolution than the diffractive structure function \FDALL, 
which is only sensitive to 
the gluon content via evolution. 
\FLDALL also probes
contributions from higher twists, similarly to the inclusive case, see \cite{Motyka:2012ty}. Thus, its measurement gives the opportunity of constraining the gluon contribution to diffraction in DIS and the dynamics beyond linear evolution driving this kind of interaction.

\FLDALL is a very poorly known quantity, with the only existing experimental study done by the H1 Collaboration~\cite{Aaron:2012zz} at HERA. On top of the intrinsic difficulty of disentangling diffractive from inclusive events, measuring the longitudinal structure function  requires 
variation of the centre-of-mass
energy of the lepton-hadron collisions. At HERA the former was determined by the LRG method. For the latter, four energies of the proton beam were employed, additionally selecting those events with high inelasticity of the electron (see below) where the contribution of the longitudinal structure function to the total reduced cross section is largest.
This region is difficult experimentally, since it is associated with low electron energies and with the hadronic final state being produced in the same backward pseudorapidity region as the scattered electron. 

Planned  DIS colliders like the Electron Ion Collider (EIC)~\cite{Accardi:2012qut,AbdulKhalek:2021gbh}, the Large Hadron-electron Collider (LHeC)~\cite{AbelleiraFernandez:2012cc,LHeC:2020van} or the Future-Circular-Collider in its electron-hadron option (FCC-eh)~\cite{FCC:2018byv,FCC:2018vvp} will benefit from larger integrated luminosities exceeding those at HERA by factors ${\cal O}(1000)$ and new detector techniques providing enhanced possibilities for separating diffractive from non-diffractive events.

In previous works we analysed the potential of the measurements of the diffractive  reduced cross section at the LHeC and FCC-he and at the EIC~\cite{Armesto:2019gxy,Slominski:2021zit} as well as the possibility for constraining the diffractive parton distribution functions. In this work we focus on the possibilities for the determination of \FLDALL in coherent diffraction on protons at the EIC, where simulations of the forward detectors, including their effect on particle reconstruction, are available~\cite{AbdulKhalek:2021gbh}.

The manuscript is organised as follows. In Sec.~\ref{sec:kinematics} we present the general expressions and kinematics that will be used in our analysis and discuss the experimental aspects of proton tagging at the EIC. In Sec.~\ref{sec:setup} we discuss the generation of 
simulated EIC data (`pseudodata')  
and the method of extraction of \FLDALL  as well as the choices of beam energies. 
Results are presented in Sec.~\ref{sec:results}, first for the reduced cross section \srD and 
then for \FLD. We then proceed to discuss the influence of the systematic error assumed in the pseudodata and the assumptions on the beam 
configurations.
Results for $R^\mathrm{D(3)}=\FLD/\FTD$ are also presented. We end with conclusions in Sec.~\ref{sec:conclu}.%\naout{,

\section{Definitions and kinematics}
\label{sec:kinematics}

\subsection{Diffractive variables and definitions}
\label{subsec:variables}

\begin{figure}[htb]
\centerline{%\fbox{
% height=0.8\columnwidth,
\includegraphics[width=0.7\columnwidth, clip]{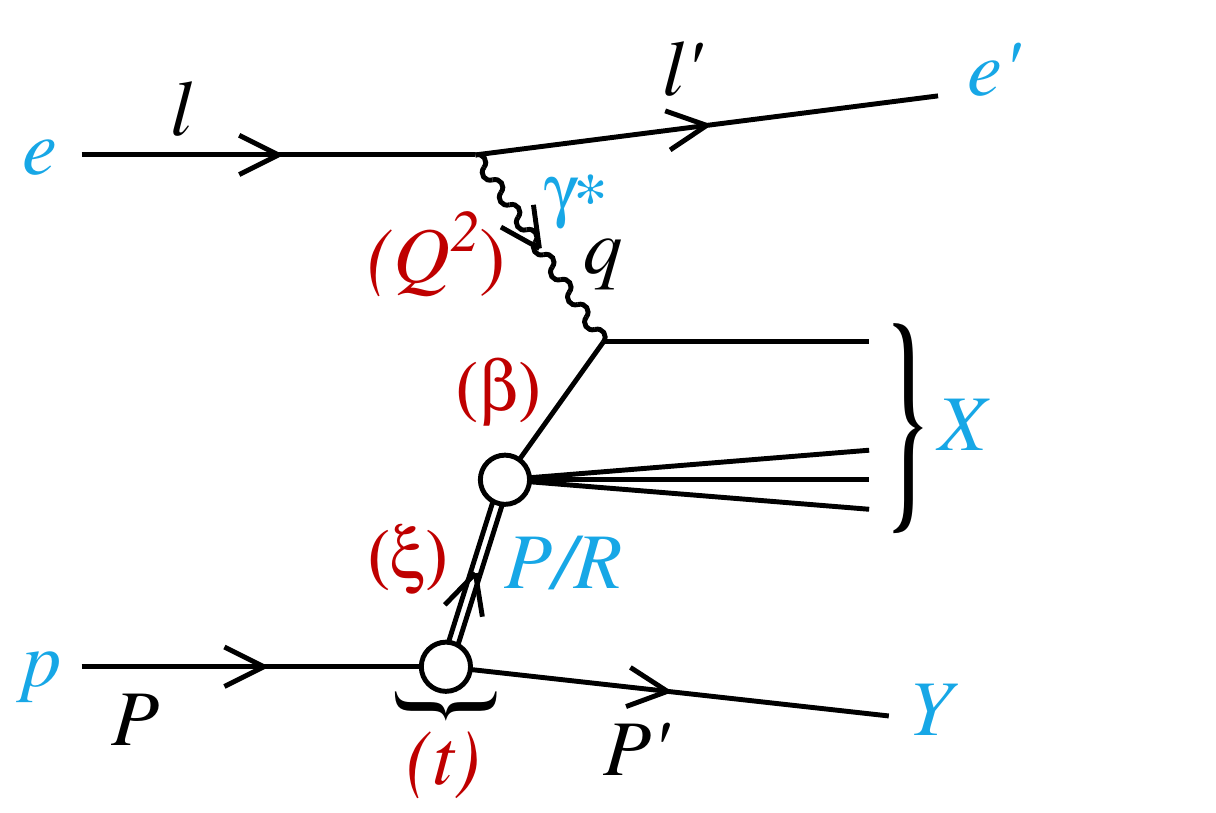}
}%}
\caption{Diagram showing the neutral current diffractive DIS process and the relevant kinematic variables in the one photon exchange approximation.}
\label{fig:kinvar}
\end{figure}

In this work we focus on neutral current diffractive deep inelastic scattering (DDIS) in the one photon exchange approximation, neglecting radiative corrections whose contribution can be corrected.
For an electron or positron with four momentum $l$ and a proton with four-momentum $P$, the diagram is shown in Fig.~\ref{fig:kinvar}.
A characteristic feature of the diffractive process, as illustrated in Fig.~\ref{fig:kinvar}, is the presence of the rapidity gap between the final proton (or its dissociated state) $Y$ and the system $X$. It is mediated by the colourless object, indicated by $P/R$, to which we refer generally as `diffractive exchange'.

    % +++++++++++++++++++++++++++++++++++++++++++++++++++++++++++++++++++

\begin{figure}[htb]
\centerline{%\fbox{
% height=0.8\columnwidth,
\includegraphics[width=0.7\columnwidth, clip]{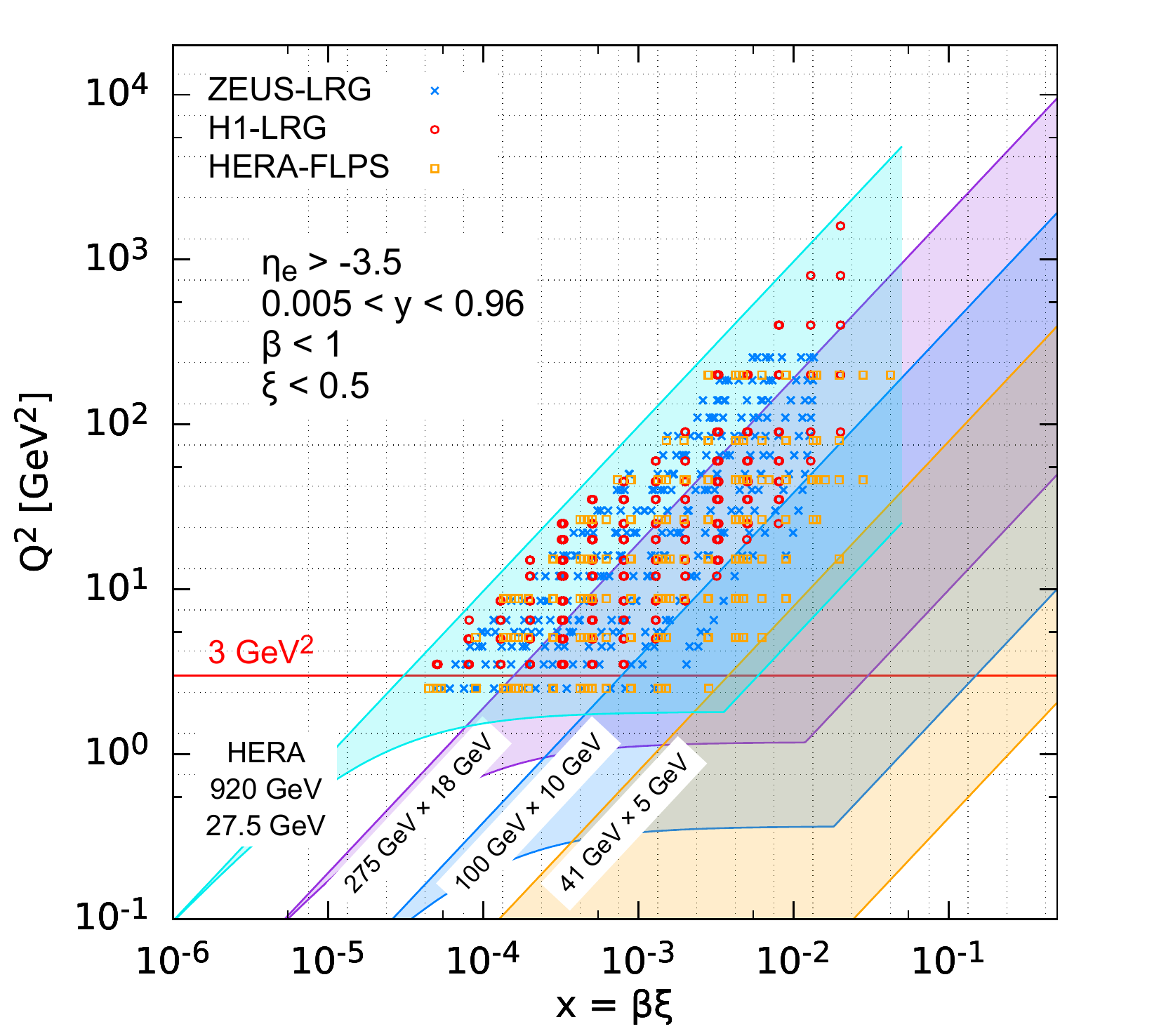}
}%}
 \caption{Kinematic $x-Q^2$ plane showing different choices of beam energies at the EIC and the region covered by HERA experiments. Note that $\eta_e > -3.5$ corresponds to an angular acceptance of 176.5 degrees for the electron.
}
\label{fig:kin}
\end{figure}

In DDIS several variables can be defined
in terms of the four-momenta indicated in 
Fig.~\ref{fig:kinvar} and the usual Mandelstam variables:
\begin{eqnarray}
\nonumber \\
Q^2 &=& -q^2\ , \nonumber \\
y &=& \frac{P\cdot q}{P\cdot \ell}\ , \nonumber \\
x &=& \frac {Q^2}{2 P\cdot q}
    = \frac{Q^2}{y s}
    \ ,\nonumber \\
\beta &=& \frac{Q^2}{2\, (P-P^\prime) \cdot q}\ ,
\nonumber \\
\xi &=& \frac{x}{\beta}\ ,\nonumber \\
t&=&(P^\prime-P)^2\ .
\end{eqnarray}
 Besides the standard DIS variables $s,Q^2,y,x$, in DDIS some additional variables appear: $t$ is the squared
 four-momentum transfer   at the proton vertex, $\xi$ (alternatively denoted by $x_\pom$)  can be interpreted as  the momentum fraction of the `diffractive exchange'   with respect to the beam hadron, and  $\beta$ 
is the momentum fraction of the parton (probed by the virtual photon) with respect to the diffractive exchange. 
In Fig.~\ref{fig:kin} we show the kinematic coverage in $x$ and $Q^2$ of the EIC for three selected energies compared to that of HERA. Since HERA was operating at higher centre-of-mass energy than the EIC, it could reach lower  values of $x$. The EIC can operate at several energy combinations, which will result in a wide coverage of $x$ also towards moderate and large $x$, and which is essential for \FLDALL measurement. In Fig.~\ref{fig:kin} only three beam energy combinations are shown, a subset of a wider range of combinations possible at the EIC, see the discussion below.  

% +++++++++++++++++++++++++++++++++++++++++++++++++++++++++++++++++++
% --- Kinematic plane

Only four variables, usually chosen to be $\beta,\xi,Q^2,t$, are needed to characterise the reduced cross section, related to the measured cross section by
\be
\frac{d^4 \sigma^{\DD}}{d\xi d\beta dQ^2 dt} = \frac{2\pi \alpha_{\rm em}^2}{\beta Q^4} \, Y_+ \, \sred^{\DD(4)}\,  ,
\label{eq:sigmared4}
\ee
where
$Y_+= 1+(1-y)^2$.
It is also customary to perform an integration over $t$, defining
\be
\label{eq:sigmared3}
\frac{d^3 \sigma^{\DD}}{d\xi d\beta dQ^2} = \frac{2\pi \alpha_{\rm em}^2}{\beta Q^4} \, Y_+ \, \sred^{\DD(3)}\, .
\ee

In the one photon exchange approximation, the reduced cross sections can be expressed in terms of two diffractive structure functions
$F_2^{\DD}$ and $F_\mathrm{L}^{\DD}$:
\be
\sred^{\DD(4)} = F_2^{\DD(4)}(\beta,\xi,Q^2,t) - \frac{y^2}{Y_+} F_\mathrm{L}^{\DD(4)}(\beta,\xi,Q^2,t) \; ,
\label{eq:sred4}
\ee
\be
\label{eq:sred3}
\sred^{\DD(3)} = F_2^{\DD(3)}(\beta,\xi,Q^2) - \frac{y^2}{Y_+} F_\mathrm{L}^{\DD(3)}(\beta,\xi,Q^2) \; ,
\ee
where $F_\mathrm{2,L}^{\DD(4)}$ have dimension $\GeV^{-2}$ and $F_\mathrm{2,L}^{\DD(3)}$ are dimensionless.

The dependence of the reduced cross sections $\sred^{\DD(4,3)}$ on the centre-of-mass 
energy comes via the inelasticity $y = \frac{Q^2}{\xi\beta s}$. Due to the $Y_+$ factor, $\sred^{\DD(4,3)} \simeq F_2^{\DD(4,3)}$ when $y$ is not too close to unity.

Both reduced cross sections $\sred^{\DD(3)}$ and $\sred^{\DD(4)}$ have been measured at HERA~\cite{Adloff:1997sc,Breitweg:1997aa,Chekanov:2005vv, Aktas:2006hx, Aktas:2006hy,Chekanov:2008fh,Chekanov:2009aa,Aaron:2010aa,Aaron:2012ad}. These data have been used for perturbative QCD analyses based on collinear factorization~\cite{Collins:1997sr,Berera:1995fj,Trentadue:1993ka}, where
the diffractive cross section reads
\be
d\sigma^{ep\rightarrow eXY}(\beta,\xi,Q^2,t) \; = \; \sum_i \int_{\beta}^{1} dz \ d\hat{\sigma}^{ei}\left(\frac{\beta}{z},Q^2\right) \, f_i^{\rm D}(z,\xi,Q^2,t) \; ,
\label{eq:collfac}
\ee
up to terms of  order  ${\cal O}(1/Q^2)$. Here,
the sum is performed over all parton species (gluon and all quark flavours).
The hard scattering partonic cross section $d\hat{\sigma}^{ei}$ can be computed perturbatively in QCD and is the same as in the inclusive deep inelastic scattering case. The long distance 
part $f_i^{\rm D}$ corresponds to the DPDFs,
which can be interpreted as conditional probabilities for partons in the proton, provided the proton is scattered into the final state system $Y$ with  four-momentum $P^\prime$. 
They are non-perturbative objects to be extracted from data, but their evolution through the DGLAP evolution equations~\cite{Gribov:1972rt,Gribov:1972ri,Altarelli:1977zs,Dokshitzer:1977sg} can be computed perturbatively,
similarly to the inclusive case.
The analogous formula for the $t$-integrated structure functions reads
\begin{equation}
\label{eq:FD3-fac}
F_{2/\mathrm{L}}^{\DD(3)}(\beta,\xi, Q^2) =
\sum_i \int_{\beta}^1 \frac{dz}{z}\,
	 C_{2/\mathrm{L},i}\Big(\frac{\beta}{z}\Big)\, f_i^{\DD(3)}(z,\xi,Q^2) \; ,
\end{equation}
where the coefficient functions $C_{2/\mathrm{L},i}$ are the same as in  inclusive DIS and the DPDFs $f_i^{\DD(3)}(z,\xi,Q^2)$ have been determined from comparisons to HERA data~\cite{Adloff:1997sc,Breitweg:1997aa,Chekanov:2005vv, Aktas:2006hx, Aktas:2006hy,Chekanov:2008fh,Chekanov:2009aa,Aaron:2010aa,Aaron:2012ad}.

\subsection{Experimental Considerations}
\label{subsec:ptag}

As can be inferred from Eq.~\ref{eq:sred3}, 
sensitivity to \FLDALL
is strongest as $y \rightarrow 1$. Experimentally, this is a region in
which backgrounds are hard to control,
since it corresponds to the lowest scattered electron energies and also
to cases where hadronic final state particles are produced in the same
(backward) pseudorapidity region as the scattered electron. 
Extractions of the inclusive and diffractive longitudinal structure
functions therefore place strong challenges on the performance of electromagnetic calorimetry, tracking and particle identification in the
backward region of the detector. H1 achieved measurements down to
electron energies of around 3~GeV. At the EIC, where charged pion rejection
factors relative to electrons of the order of $10^{-4}$ are targeted,
the aim is to go substantially lower, even into the 
sub-GeV range. Here, we apply an upper $y$ cut of 0.96,
which is typical of current EIC studies. The targeted $\eta$ range of the
EIC experiments, with calorimeter and tracking coverage to at least as 
far backwards as $\eta = -3.5$, provides full coverage for scattered
electrons with $Q^2 > 1$ GeV$^2$.

\begin{figure}[htb]
\centerline{%\fbox{
% height=0.8\columnwidth,
\includegraphics[width=0.9\columnwidth,clip,trim=0 0 0 25]{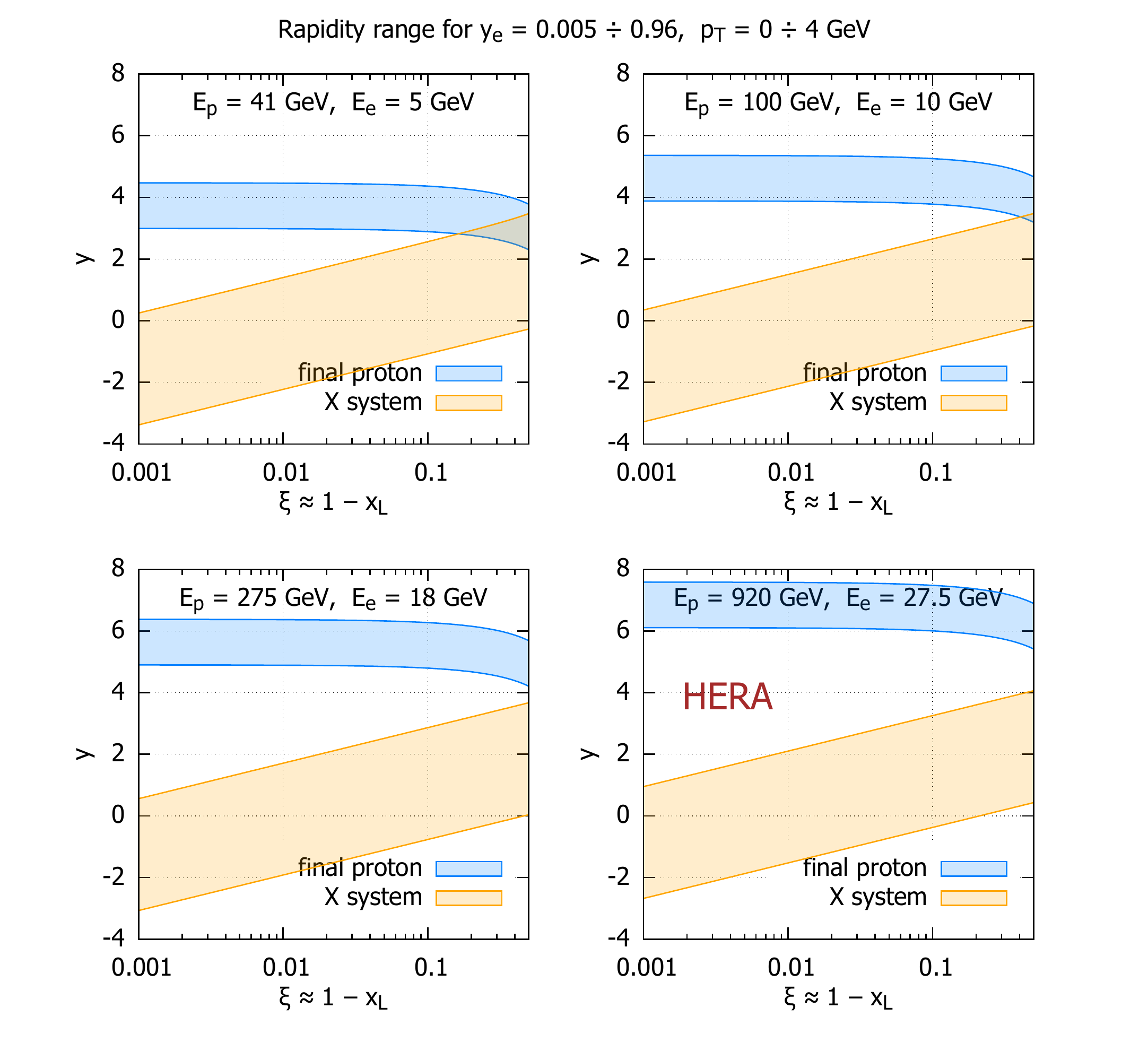}
}%}
\caption{Ranges in the rapidity of the scattered proton and 
the undecayed 
dissociative 
system $X$ as a function of $\xi$ 
for three different beam energy combinations at the EIC
and for HERA. The bands correspond to all cases where the proton
transverse momentum is lower than 4 GeV, $0.005<y<0.96$  and
$0.1 < \beta <0.9$.}
\label{fig:rapgap}
\end{figure}

In the H1 measurement~\cite{Aaron:2012zz}, the use of the LRG method
of selecting diffractive events led to a normalisation uncertainty
of 7\%. This uncertainty can in principle be eliminated through the
use of beamline proton tagging based on
instrumentation housed in Roman pot insertions to the beampipe. 
In contrast to the situation at previous colliders, beamline instrumentation has been a fundamental consideration from the outset
at the EIC. It is essential to the success of the diffractive
programme, as illustrated in~\Fig{fig:rapgap}, where the 
rapidity ranges covered by the undecayed final state system $X$
and proton are shown as a function of $\xi$ for
four combinations of electron and proton beam energies.
The bands correspond to ranges \(\beta \in [0.1, 0.9]\)
and \(p_\perp\) of the final state proton below \(4\, \GeV\).
The
decay of the $X$ system into a multi-particle hadronic system 
extends its extent forwards in pseudorapidity in a manner that scales logarithmically
with $\xi$, reducing the size the rapidity gap.
%It should be kept in mind that 
For comparison, gaps smaller than about 
3 pseudorapidity units could not be used reliably at HERA 
due to
the poorly modelled contributions from 
gaps produced from hadronisation fluctuations in non-diffractive processes. 
Whilst fairly large rapidity gaps exist at the lowest $\xi$
values and highest EIC centre-of-mass energies, it is clear that
throughout most of the EIC phase space and for most of the 
expected beam configurations, LRG methods will yield poor
performance. 

\begin{figure}[hbt!]
\centerline{%
\includegraphics[width=\columnwidth,clip]{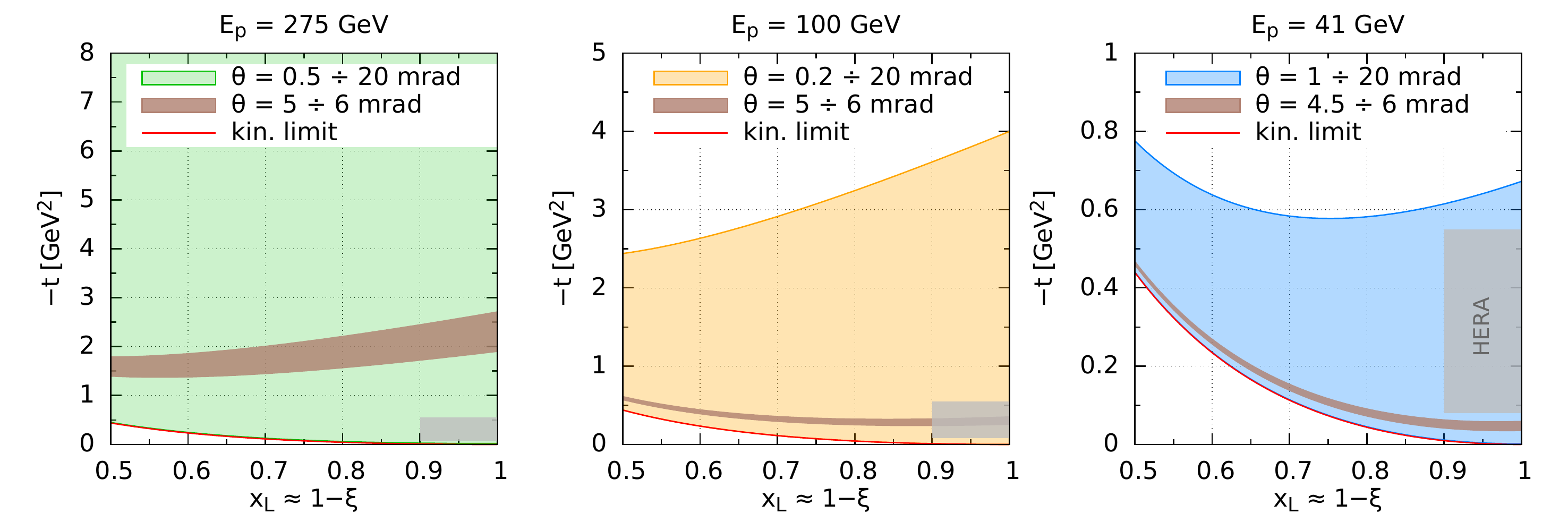}%
}
\caption{Final proton tagging.
 \(\xL, t\) range of the proton tagged by the EIC detector for three proton energies, 275 GeV, 100 GeV and 41 GeV.
The brown strip marks a small ($\sim$ 1 mrad) region not covered by the current detector design.}
\label{fig:xL-t}
\end{figure}

The most recent studies of the physics
and detector requirements at the EIC
envisage multiple beamline proton spectrometers,
allowing a full determination of the outgoing
proton kinematics with good measurements of both
$\xi$ and $t$. In Fig.~\ref{fig:xL-t} we show the 
kinematic coverage for the forward proton considered in~\cite{AbdulKhalek:2021gbh}. 
In contrast to the LRG method,
the multiple planned detector stations with a combined angular 
acceptance $0.5-20$ mrad lead to a wide potential
measurement range in $\xi$ and $t$
for all beam energies. 
Measurements up to $\xi$
values as large as 0.5 may be possible, well beyond the range 
in which diffractive ($t$-channel exchange) processes
are expected to be the dominant mechanism 
for leading proton production. We therefore 
assume here that events will be selected based on the scattered
proton and that the accessible phase space in $\xi$ is not 
strongly limited experimentally.

The standard Rosenbluth method of extracting longitudinal structure
functions
involves fits to data at the same
$\xi$, $\beta$
and $Q^2$ values, but
different centre-of-mass energies. Systematic uncertainties
that are not correlated between different beam energies 
therefore tend to propagate
into larger uncertainties on \FLDALL than those
that are positively correlated between 
beam energies.
Since statistical uncertainties are completely uncorrelated between different
beam energies, \FLDALL measurements are also particularly
sensitive to the available sample sizes. 

Due to the relatively small integrated luminosities
in the reduced
proton beam energy runs, 
the HERA measurement of \FLD~\cite{Aaron:2012zz}
was limited by statistical uncertainties
throughout most of the phase space. 
Since the integrated luminosity expected at the EIC is around 
three orders of magnitude larger than that at HERA, the
sample sizes will be much larger
(integrated luminosities of 10 fb$^{-1}$ per beam energy are assumed here)
and statistical uncertainties are expected to be unimportant. 

A detailed systematic uncertainty analysis was carried out
in the HERA measurement, with the conclusion that no single source
dominated, but also giving some baseline from which to extrapolate
to the 
likely precision achievable on the cross sections
at a future collider such as the EIC. 
The best
precision achieved in
diffractive reduced cross section measurements at
HERA was at the 4\% level, with uncorrelated sources
contributing as little as 2\%, arising primarily from
track-cluster linking and vertex finding efficiencies.
Recent and ongoing studies of proposed
EIC instrumentation solutions \cite{AbdulKhalek:2021gbh}
already indicate that uncertainties of this kind will be
dramatically reduced at the EIC. We therefore consider
scenarios in which the uncertainties that are uncorrelated
between beam energies are either 1\% or 2\%. 
With sources related to the LRG method eliminated, correlated
systematic 
uncertainties are also expected to be reduced significantly.
The alignment and calibration procedures required in
Roman pot methods inherently lead to correlated systematics,
but using methods developed at 
HERA \cite{ZEUS:1996bgb,VanEsch:1999pi,Astvatsatourov:2014dna},
coupled with the substantial
further evolution of proton-tagging techniques at the 
LHC \cite{AbdelKhalek:2016tiv,Adamczyk:2017378,TOTEM:2017asr,CMS:2018uvs} and future EIC-specific work, we estimate that these are controllable to 
the sub-2\% level, and will thus 
have a negligible effect on the \FLDALL extraction compared with
the uncorrelated sources.

%%%%%%%%%%%%%%%%%%%%%%%%%%%%%%%%%%%%%%%%%%%%
\section{Method}
\label{sec:setup}

\subsection{Pseudodata generation}
\label{subsec:pseudodata}

We shall first describe the pseudodata generation for our simulations. The momentum transfer $t$ is integrated over in this analysis.
Let us rewrite Eq.~(\ref{eq:sred3}) as
\be
\label{eq:sred3_1}
\sred^{\DD(3)}
=  F_2^{\DD(3)}(\beta,\xi,Q^2) - \YL \, F_\mathrm{L}^{\DD(3)}(\beta,\xi,Q^2)\; ,
\ee
where
\be
\label{eq:YL}
\YL = \frac{y^2}{Y_+} = \frac{y^2}{1+(1-y)^2} \; .
\ee
    As mentioned previously, the extraction of the longitudinal diffractive structure function relies on the possibility of  disentangling it from $F_2^D$, as is evident in the formula above for the reduced cross section. This is possible if, for fixed $(\beta,Q^2,\xi)$, one can vary $\YL$,
    and hence $y$, in a sufficiently wide range. 
    Given that $y=Q^2/(s\beta\xi)$
    it is therefore necessary to perform measurements of the reduced cross section using different centre-of-mass energies. The EIC is uniquely positioned to perform such a measurement, thanks to its design, which allows for a wide range of different beam energies.

    We have considered several beam energies for both 
    the electrons and the protons, within the range
    expected for the EIC:
    \begin{align} 
 E_e &= 5, 10, 18 \, \GeV\, , \nonumber \\
E_p &= 41, 100, 120, 165, 180, 275 \, \GeV \, .
\end{align}
   
These beam energies combine to give 17 distinct centre-of-mass energies (there is a degeneracy in this  choice since two combinations $10\times 180$ and $18\times 100$ lead to the same centre-of-mass energy, $85 \, \GeV$).   
The centre-of-mass energies corresponding to all combinations
are given in Table~\ref{tbl:ecm}.
In order to test the sensitivity of \FLD to the 
available beam energies,
we consider three 
different subsets in the analysis :
\begin{itemize}
\vspace{-1em}
\setlength\itemsep{0.1em}
\item [\EsetAll)] 17 values --- all combinations from Table \ref{tbl:ecm} except for 
$10\times 180$. 
\item [\EsetBig)] 9 values --- marked bold in Table \ref{tbl:ecm},
\item [\EsetMin)]  5 values --- marked bold against a green background in Table \ref{tbl:ecm}.
\end{itemize}
Set S-17 contains the widest range of possibilities.
S-5 is the set of combinations that has often been assumed in EIC studies to date \cite{AbdulKhalek:2021gbh}. Additionally, we consider an intermediate set S-9, which restricts the list to
three proton and three electron beam energies,
whilst maintaining the same overall kinematic range as \EsetAll.

\renewcommand{\arraystretch}{1.3}
\begin{table}[htbp]
\begin{center}
\begin{tabular}{cr|*6r}
	& & \multicolumn{6}{c}{\(E_p\, [\GeV]\)} \\
 & & 41 & 100 & 120 & 165 & 180 & 275 \\
 \hline
 \multirow{3}{*}{\rotatebox{90}{\(E_e\, [\GeV]\;\)}}
 & \vbox to 12pt{}
 5 & \sBD{29} & \sBD{45} & 49       & \sB{57}  & 60       & 74 \\
 & 10 & 40 & \sBD{63} & 69 & \sB{81} & 85 & \sBD{105} \\
& 18 & 54 & \sB{85} & 93 & \sB{109} & 114 & \sBD{141} \\
\end{tabular}
\end{center}
\caption{Centre-of-mass energies (in GeV) for various combinations of beam energies.}
\label{tbl:ecm}
\end{table}
\goodbreak

The pseudodata for the reduced diffractive cross section at the EIC were generated using Eqs.~\Eq{eq:sred3} and \Eq{eq:FD3-fac}. The diffractive parton distribution used for the evaluation of the cross section is the  ZEUS-SJ set~\cite{ZEUS:2009uxs}. This fit uses inclusive diffractive data together with diffractive DIS dijet data, which are added to improve the constraints on the diffractive gluon distribution.

The details of the ZEUS-SJ parametrization closely follow those
of \cite{H1:2006zyl} and can be found in~\cite{ZEUS:2009uxs}. Below we summarize a few important features. The diffractive parton densities are parametrized using a two-component form:
\be
f_i^{\DD(4)}(z,\xi,Q^2,t) =  f^p_{\pom}(\xi,t) \, f_i^{\pom}(z,Q^2)+f^p_{\regg}(\xi,t) \, f_i^{\regg}(z,Q^2) \;.
\label{eq:param_2comp}
\ee
The first term in Eq.~(\ref{eq:param_2comp}) is interpreted as the exchange of a `Pomeron' and the second is a `Reggeon' component. They dominate in different $\xi$ regions: the `Pomeron' is dominant for $\xi \le 0.01$. The `Reggeon' starts to be important for $\xi > 0.01$ and becomes dominant for $x > 0.1$.
For both terms,
proton vertex factorization is assumed, which means  that the diffractive parton density factorizes into a parton distribution in a diffractive exchange $f_i^{\pom,\regg}$ and a flux factor $f^p_{\pom,\regg}$. The parton distribution in  the  `Pomeron' and `Reggeon'  $f_i^{\pom,\regg}(\beta,Q^2)$ only depend on the longitudinal momentum fraction $\beta$ of the parton with respect to the Pomeron/Reggeon and the photon virtuality $Q^2$. The flux factors $f^p_{\pom,\regg}(\xi,t)$, on the other hand, only depend on $\xi$, which is related to the size of the rapidity gap, and the momentum transfer at the proton vertex $t$. They represent the probability that a Pomeron/Reggeon with given values 
of $\xi,t$ couples to the proton. 
The flux factors are parametrized using a form motivated by  Regge theory:
\be
 f^p_{\pom,\regg}(\xi,t) = A_{\pom,\regg} \frac{e^{B_{\pom,\regg}t}}{\xi^{2\alpha_{\pom,\regg}(t)-1}} \, ,
\label{eq:flux}
\ee
with a linear trajectory ${\alpha_{\pom,\regg}(t)=\alpha_{\pom,\regg}(0)+\alpha_{\pom,\regg}'\,t}$.

The diffractive parton distributions are evolved using the 
NLO 
DGLAP equations. For the case of the Pomeron at the initial scale $\mu_0^2 = 1.8\,\GeV^2$ they are parametrized as
\be
z f_i^\pom (z,\mu_0^2)= A_i z^{B_i} (1-z)^{C_i} \; ,
\ee
where $i$ is a gluon or a light quark. In the diffractive parametrizations all the 
light quarks (anti-quarks) are assumed to be equal. 
For the treatment of heavy flavours, a variable flavour number  scheme (VFNS) is adopted, where the charm and bottom quark DPDFs are generated radiatively via DGLAP evolution. There is no intrinsic heavy quark distribution present.
The structure functions are calculated in a General-Mass Variable Flavour Number  scheme (GM-VFNS) \cite{Collins:1986mp,Thorne:2008xf} which 
ensures a smooth transition of $F_\mathrm{2,L}$ across the flavour thresholds by including
$\mathcal{O}(m_h^2/Q^2)$ corrections.

The model that we use to compute the diffractive cross section is state of the art. Still, it contains substantial uncertainties in 
the Reggeon contribution, which was poorly constrained by HERA data. 
Whilst these uncertainties affect the predicted cross sections and 
structure functions, they do not to first order impact 
our assessment of the feasibility of measuring the longitudinal diffractive structure function, which is our primary objective here.

%%%%%%

The parton distributions for the Reggeon component are taken from a parametrization which was obtained from fits to the pion structure function \cite{Owens:1984zj,Gluck:1991ey}.
HERA data required the addition of the Reggeon contribution, but could not constrain it. The high $\xi$ region where it dominates is accessible in the EIC kinematics, and the possibilities for disentangling the Reggeon contribution (or any contribution other than the Pomeron) were discussed in~\cite{AbdulKhalek:2021gbh}. This is an aspect demanding a dedicated study that we leave for the future.

The pseudodata were generated as the extrapolation of the fit to HERA~\cite{ZEUS:2009uxs}, 
amended with a random Gaussian smearing 
with  standard deviation corresponding to the relative 
error $\delta$.  The total error was assumed to be composed of systematic and statistical components and computed as
\be
\delta = \sqrt{\delta^2_{\rm sys}+\delta^2_{\rm stat}}\, .
\ee
The statistical error was evaluated assuming an  integrated luminosity $10 \, {\rm fb}^{-1}$, see~\cite{AbdulKhalek:2021gbh}.
For the binning adopted in this study, the statistical uncertainties have a very small effect on the extraction of the longitudinal structure function.
As discussed in Sec.~\ref{subsec:ptag}, correlated systematic uncertainties on the reduced cross section are also expected to be 
relatively unimportant in the \FLDALL extraction
and are thus neglected here.
For the uncorrelated systematic error we have considered two scenarios, with $2\%$ and $1\%$ ascribed to each data point.

The cuts imposed for the data selection are:
\begin{itemize}
    \item $Q^2 \ge 3 \; \GeV^2$ :  
    both the H1~\cite{Aktas:2006hy} and the ZEUS~\cite{ZEUS:2009uxs} analyses of the inclusive diffractive data observed that the 
    quality of the DGLAP-based fit deteriorates in 
the low $Q^2$ region, possibly because it is sensitive to higher twist contributions. 
This cut is imposed to limit the sensitivity to 
such effects. 
The EIC kinematics and expected
scattered electron coverage lead to full acceptance for all
$x$ in the chosen $Q^2$ region at all beam energies,
see Fig.~\ref{fig:kin}.
     
    \item
    $y$ between 0.005 and 0.96, which is the expected coverage
    of a typical measurement whilst maintaining well-controlled
    systematics. 
\end{itemize}

\begin{figure}[htb]
\centerline{%\fbox{
% height=0.8\columnwidth,
% --- trim: left bottom right top
%\includegraphics[width=0.9\columnwidth, clip,trim=0 0 0 25]{figs/sigYL2_S5_a1CL68.pdf}
%}%}
\includegraphics[width=0.9\columnwidth, clip,trim=0 0 0 40]{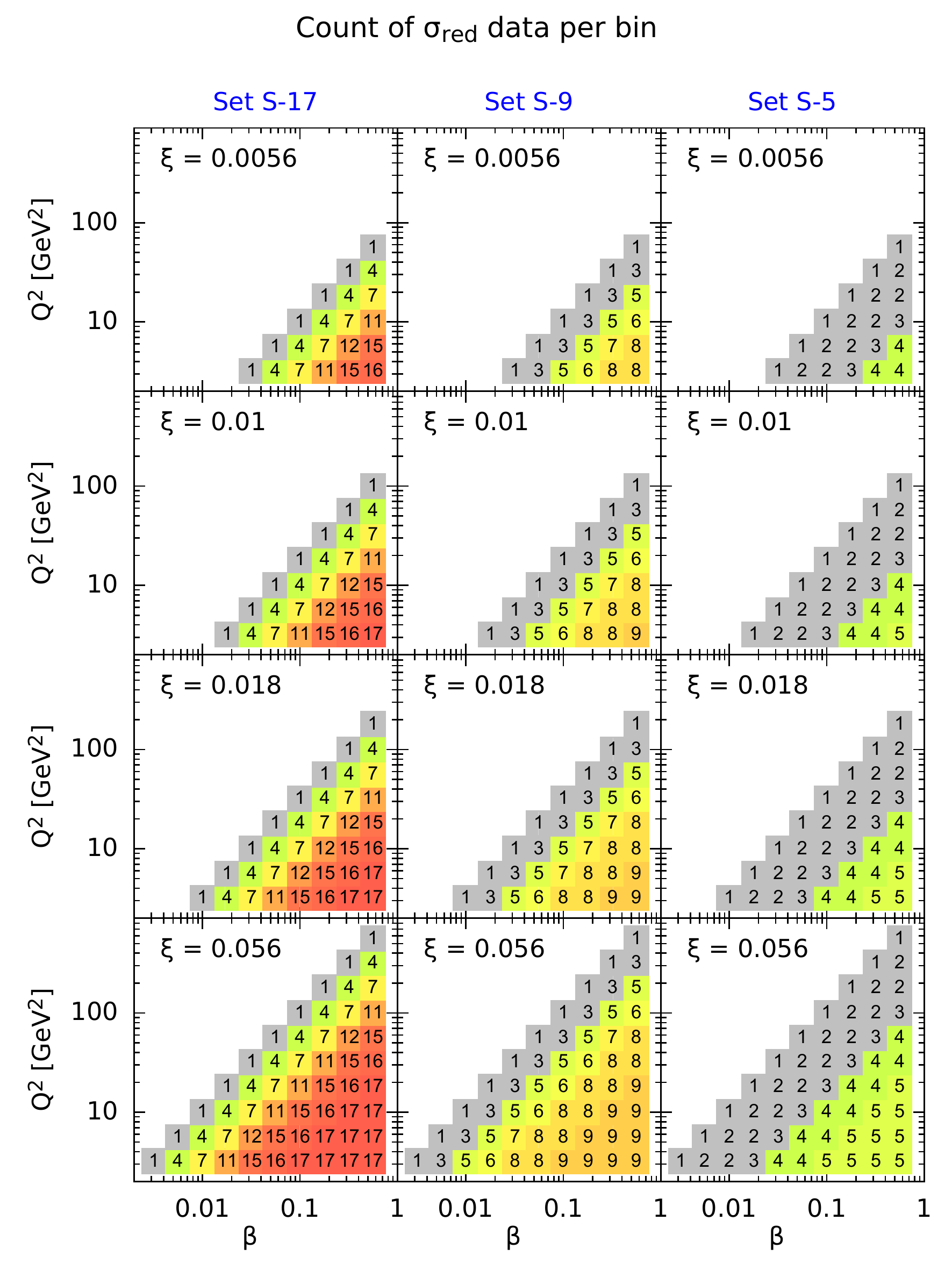}
}%}
\caption{Count of different beam energy combinations from 
among set \EsetAll that lead to measurable
$\sred^{\DD(3)}$ data points 
%vs. \YL per \
for each $(\xi, \beta,Q^2)$ bin.
Only cases with a number of counts $\ge 4$ are considered for the extraction of \FLD.}
\label{fig:YLcount}
\end{figure}

We  adopt a uniform logarithmic binning
with  four bins per decade for each of $\xi,\beta$ and $Q^2$ \footnote{Resolution studies, performed using the \textsf{Rapgap} Monte Carlo generator (\cite{Jung:1993gf}, see also {\tt https://rapgap.hepforge.org/}) with the smearing routines available for the detector model in~\cite{AbdulKhalek:2021gbh}, show that this binning is perfectly achievable.}. This results in 
the numbers of \sred data points for each \((\xi,\beta, Q^2)\) bin as shown in Fig.~\ref{fig:YLcount}. Taking into account that we require four points for the linear fit 
(see Sec.~\ref{subsec:extraction}) to extract \FLD, 
the analysis therefore proceeds with
a total of
364 \FLD values for set \EsetAll,
285 \FLD values for set \EsetBig  and  
160 \FLD values for set \EsetMin.

\subsection{Extraction of the diffractive longitudinal structure function}
\label{subsec:extraction}

\begin{figure}[htb]
\centerline{%\fbox{
% height=0.8\columnwidth,
% --- trim: left bottom right top
%\includegraphics[width=0.9\columnwidth, clip,trim=0 0 0 25]{figs/sigYL2_S5_a1CL68.pdf}
%}%}
\includegraphics[width=0.9\columnwidth, clip,trim=0 10 0 25]{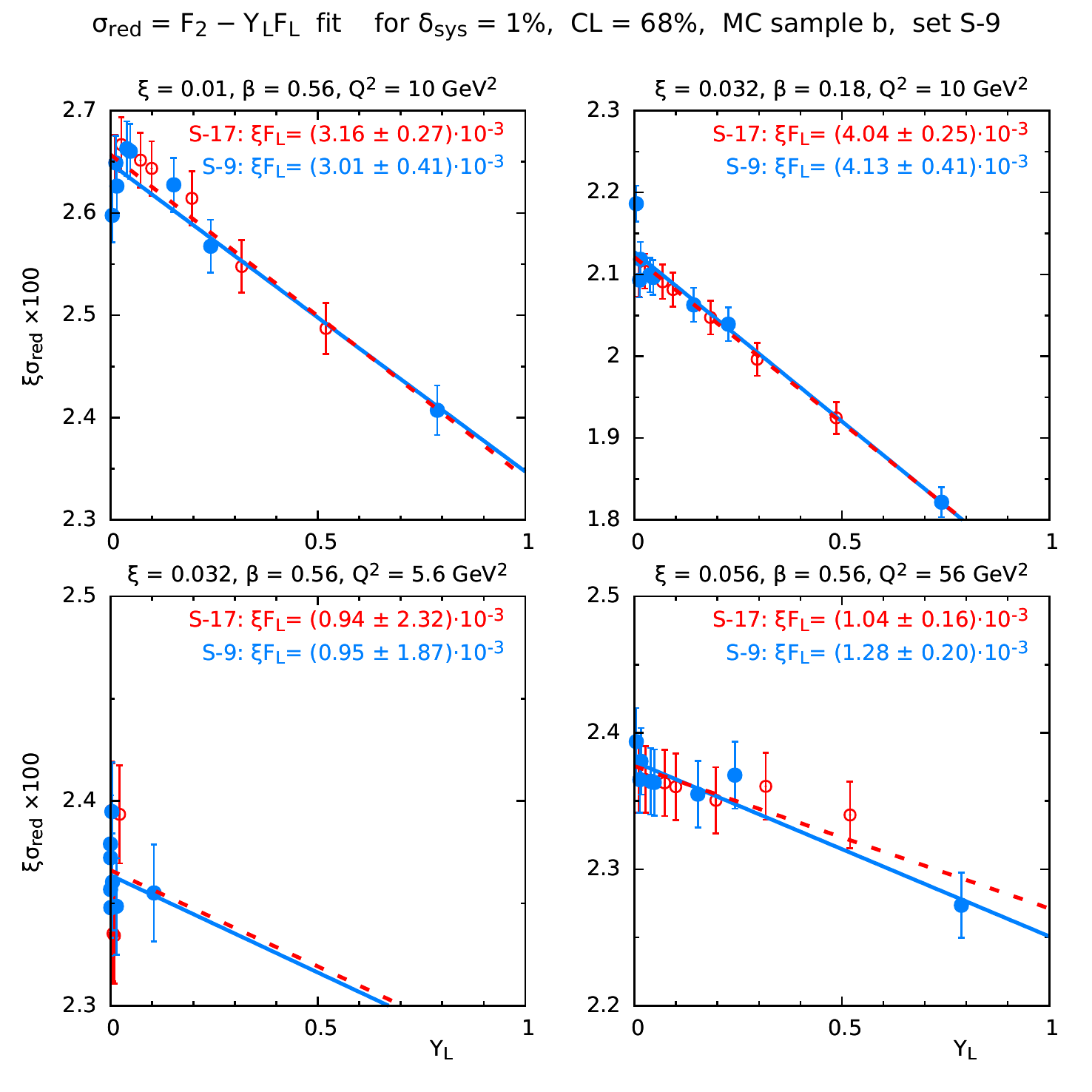}
}%}
\caption{
Examples of the $\sred^{\DD(3)}=\FtwoD - \YL \FLD$ fit. Comparison between set \EsetAll (red line and points) and \EsetBig (blue line and points). }
\label{fig:FL_fit_AB}
\end{figure}

\begin{figure}[htb]
\centerline{%\fbox{
% height=0.8\columnwidth,
%\includegraphics[width=0.9\columnwidth, clip,trim=0 0 0 25]{figs/sigYL2_S4_a1CL68.pdf}
%}%}
\includegraphics[width=0.9\columnwidth,clip,trim=0 10 0 25]{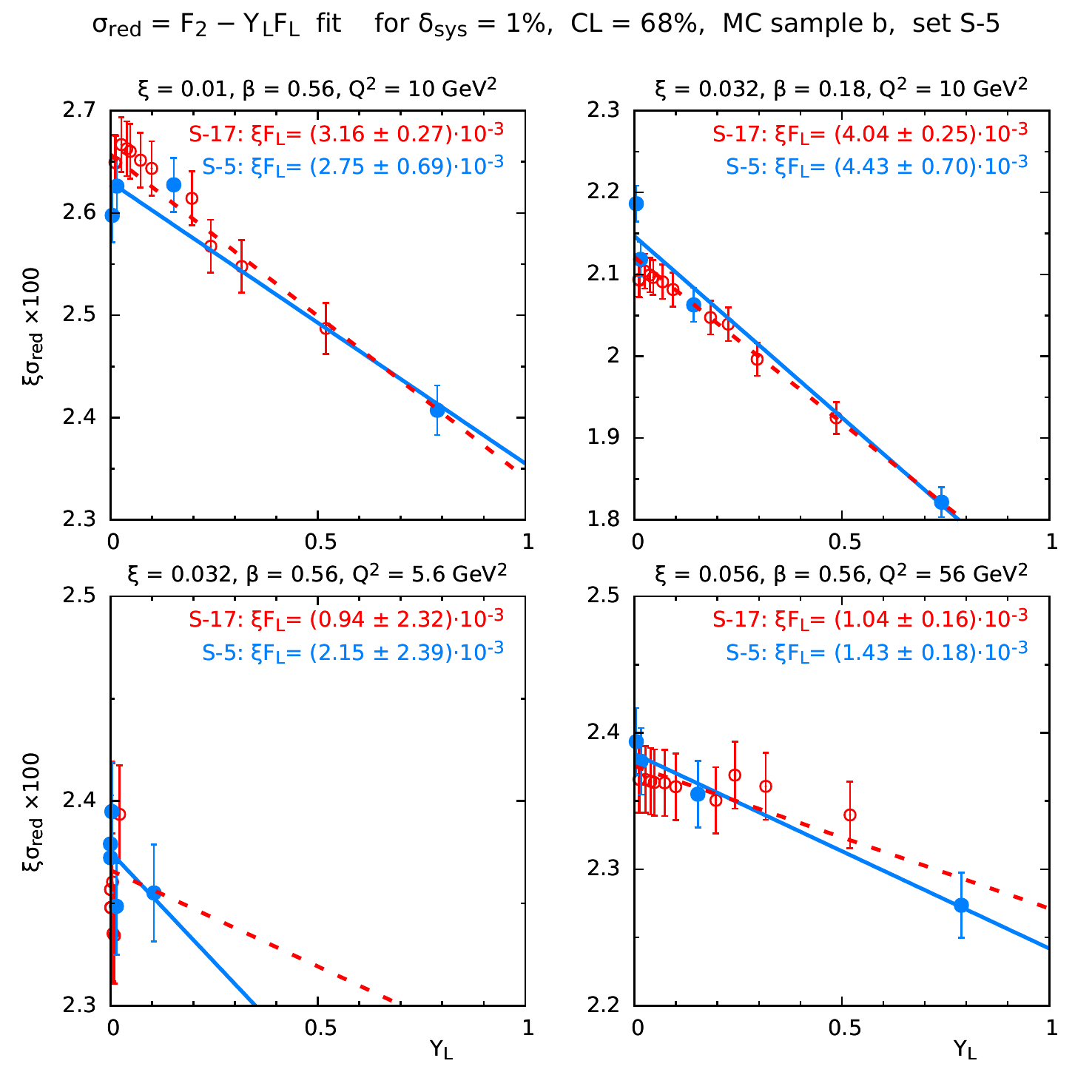}
}%}
\caption{Examples of the $\sred^{\DD(3)}=\FtwoD - \YL \FLD$ fit. Comparison between set \EsetAll (red line and points) and \EsetMin (blue line and points).
}
\label{fig:FL_fit_AD}
\end{figure}

The extraction of the  diffractive longitudinal structure function \FLD is performed using the same method as in the H1 analysis~\cite{Aaron:2012zz}.  This method was adapted from the measurements of the inclusive longitudinal structure function $F_\mathrm{L}$, see~\cite{H1:2008rkk,ZEUS:2009nwk,H1:2010fzx}. The reduced cross section is a linear function of $\YL$, see Eq.~\eqref{eq:sred3_1}. 
The structure function $\FLD$ can thus be found by performing a linear fit, and extracting the slope of $\sred^{\DD(3)}$ as a function of $\YL$.  This is done for 
every set of values  in $Q^2$, $\xi$ and $\beta$ 
for which there are four or more available \sred values (in the H1 analysis~\cite{Aaron:2012zz} only three points were 
required). In Figs.~\ref{fig:FL_fit_AB} and \ref{fig:FL_fit_AD} examples of fits in 4 bins of 
$(\xi,\beta,Q^2)$ are shown, which are discussed in detail in Subsec.~\ref{subsec:syst}.


% ===================================================================
\section{Results}
\label{sec:results}

In this Section we show the 
simulated results for $\FLD$ and analyse the influence of choices of beam energies, systematic errors and numbers of measurements.
We also extract results for $R = \FLD/\FTD$.
In obtaining uncertainties on $\FLD$ from fits to 
Eq.~\ref{eq:sred3_1}, 
we take 68\%
confidence limits (CL).
This practically corresponds to 
$1 \sigma$ errors for a number of degrees of freedom \(\mathrm{NDF} \gtrsim 10\).
However, many \((\xi, Q^2, \beta)\) bins where \FLD is fitted contain even as few data points as 4,
which results in 68\%
CL error equal to  \(\simeq 1.3
\, \sigma\).

  % +++++++++++++++++++++++++++++++++++++++++++++++++++++++++++++++++++

% +++++++++++++++++++++++++++++++++++++++++++++++++++++++++++++++++++

\subsection{Influence of systematic errors and choices of beam energies}
\label{subsec:syst}

In Figs.~\ref{fig:FL_fit_AB} and \ref{fig:FL_fit_AD} 
examples of fits are shown in 4 selected bins of 
$(\xi,\beta,Q^2)$. In each figure two data sets are shown. The open red circles correspond to set \EsetAll, and the filled 
blue points 
are the subset that is also present in \EsetBig (for Fig.~\ref{fig:FL_fit_AB}) and \EsetMin (for Fig.~\ref{fig:FL_fit_AD}). 
Uncorrelated systematic uncertainties are considered at the level of $1\%$ on each data point, with 
the influence of correlated sources taken to be 
negligible as discussed previously. 
Separate fits are performed to each of the sets, 
resulting in the two lines shown on each of the plots.

The \EsetAll set of beam energies 
contains the most points in $\YL$ and therefore
by construction gives the most precise results. Reducing the number of 
beam energy combinations lowers the precision of the $\FLD$ extraction. However, we observe that the fits with \EsetBig and even
\EsetMin do not deviate strongly from those of \EsetAll in most
cases.
This is encouraging, since set \EsetMin is the current 
working hypothesis for the EIC energy combinations. 
It is also evident that the strongest variations with the 
choice of set arise in $(Q^2,\xi,\beta)$ bins where there is a limited 
range of $\YL$ available for the fit.  
For example,
the bin with $(\xi=0.032,\beta=0.56,Q^2=5.6 \, \GeV^2)$ has 
a very small
range in $\YL$, resulting in large variations between the sets and 
correspondingly large uncertainties on the extracted \FLD, despite the relatively large number of $Y_L$ points.  
The relative accuracy of the \FLD extraction also depends on 
its absolute size, which is reflected in the 
steepness
of the slope.

\begin{figure}[htb]
\renewcommand{\arraystretch}{1.9}
\begin{tabular}{cc}
\(E_\CM \textrm{ set \EsetAll}, \delta_\mathrm{sys} = 1\%, 68\%\ {\rm CL}\)
&
\(E_\CM \textrm{ set \EsetAll}, \delta_\mathrm{sys} = 2\%, 68\%\ {\rm CL}\)
\cr
\includegraphics[width=0.45\columnwidth, clip,trim=0 0 13 29]{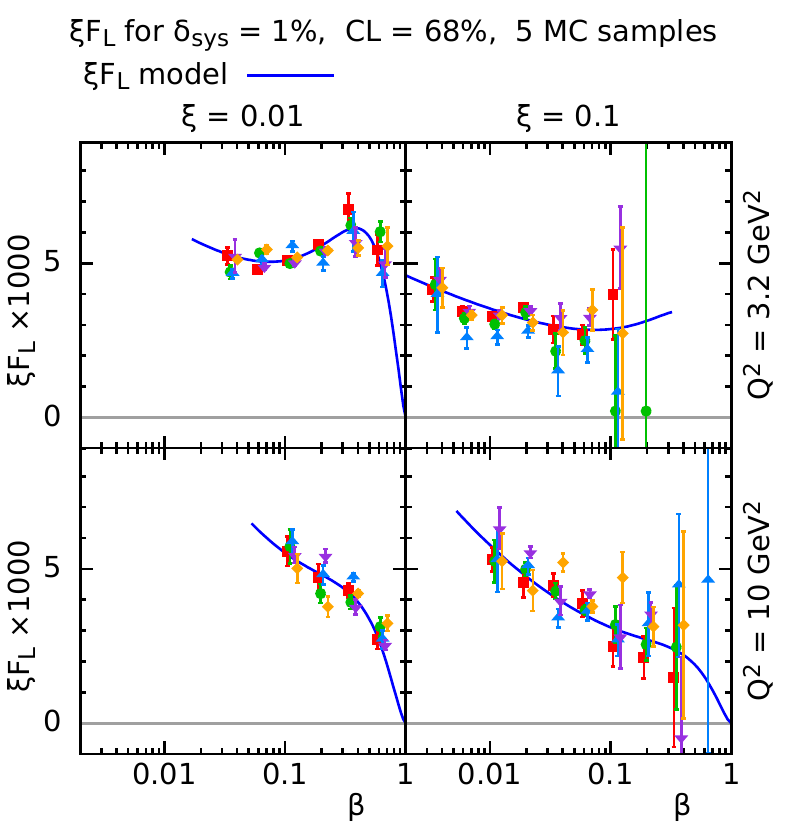}
&
\includegraphics[width=0.45\columnwidth, clip,trim=13 0 0 29]{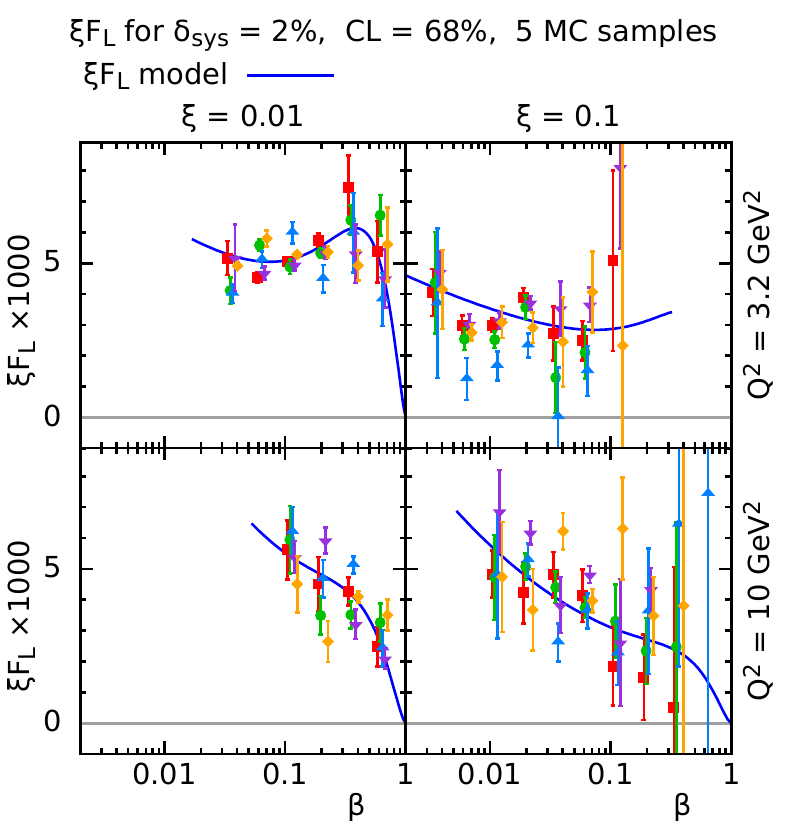}
\cr
\(E_\CM \textrm{ set \EsetBig}, \delta_\mathrm{sys} = 1\%, 68\%\ {\rm CL}\)
&
\(E_\CM \textrm{ set \EsetMin}, \delta_\mathrm{sys} = 1\%, 68\%\ {\rm CL}\)
\cr
\includegraphics[width=0.45\columnwidth, clip,trim=0 0 13 29]{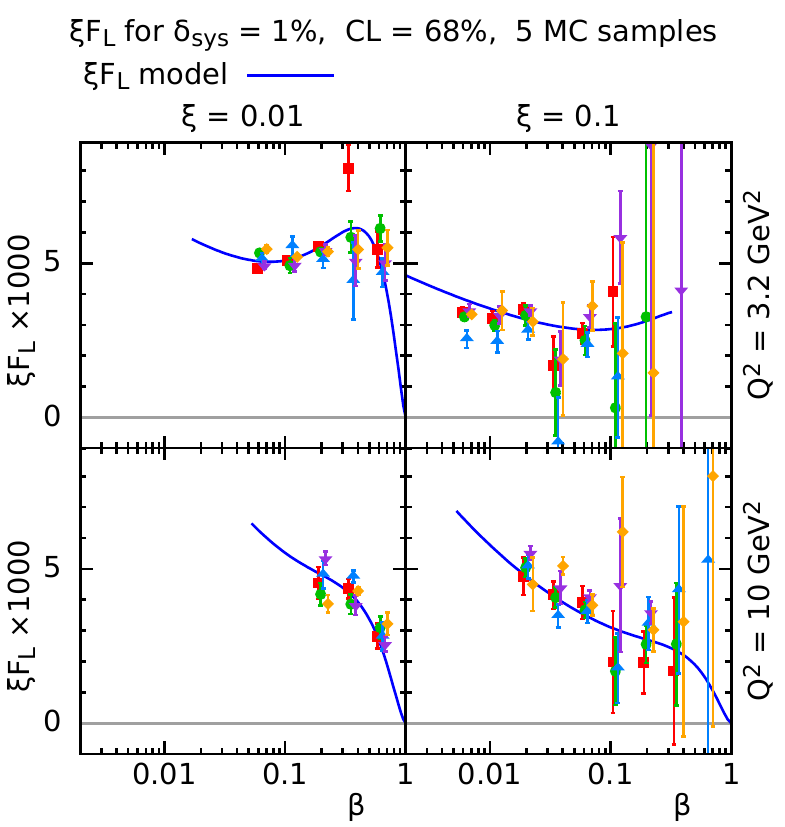}
&
\includegraphics[width=0.45\columnwidth, clip,trim=13 0 0 29]{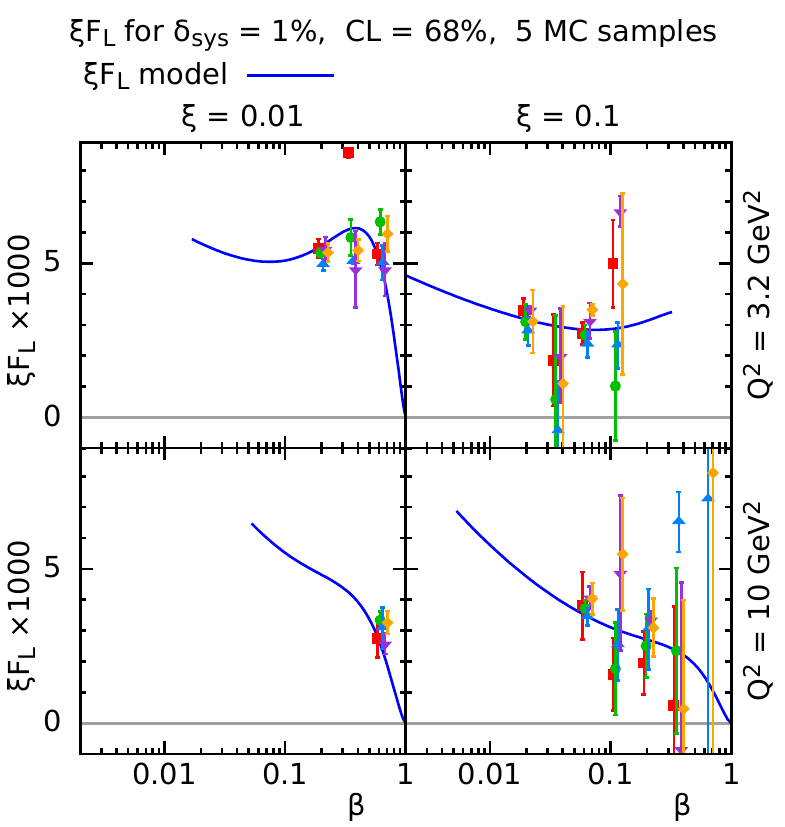}
\cr
\end{tabular}
\caption{Comparisons of the extracted \FLD in four selected $(\xi, Q^2)$ bins, for five Monte Carlo samples (markers with different colours, horizontally 
displaced from each other for clarity).
The
first row compares the results for two different \(\esys\) 
values, 1\% on the left and 2\% on the right, for the \EsetAll set of beam energies.
In the second row the results for smaller \(E_\CM\) sets (\EsetBig and \EsetMin) are shown with $\esys = 1\%$. 
The solid lines show the central values of the model used for the generation of the pseudodata from which \FLD was extracted.}
\label{fig:compar_syst_ener}
\end{figure}

\begin{figure}[htb]
\centerline{%\fbox{
% height=0.8\columnwidth,
\includegraphics[width=0.9\columnwidth,trim=0 20 0 40, clip]{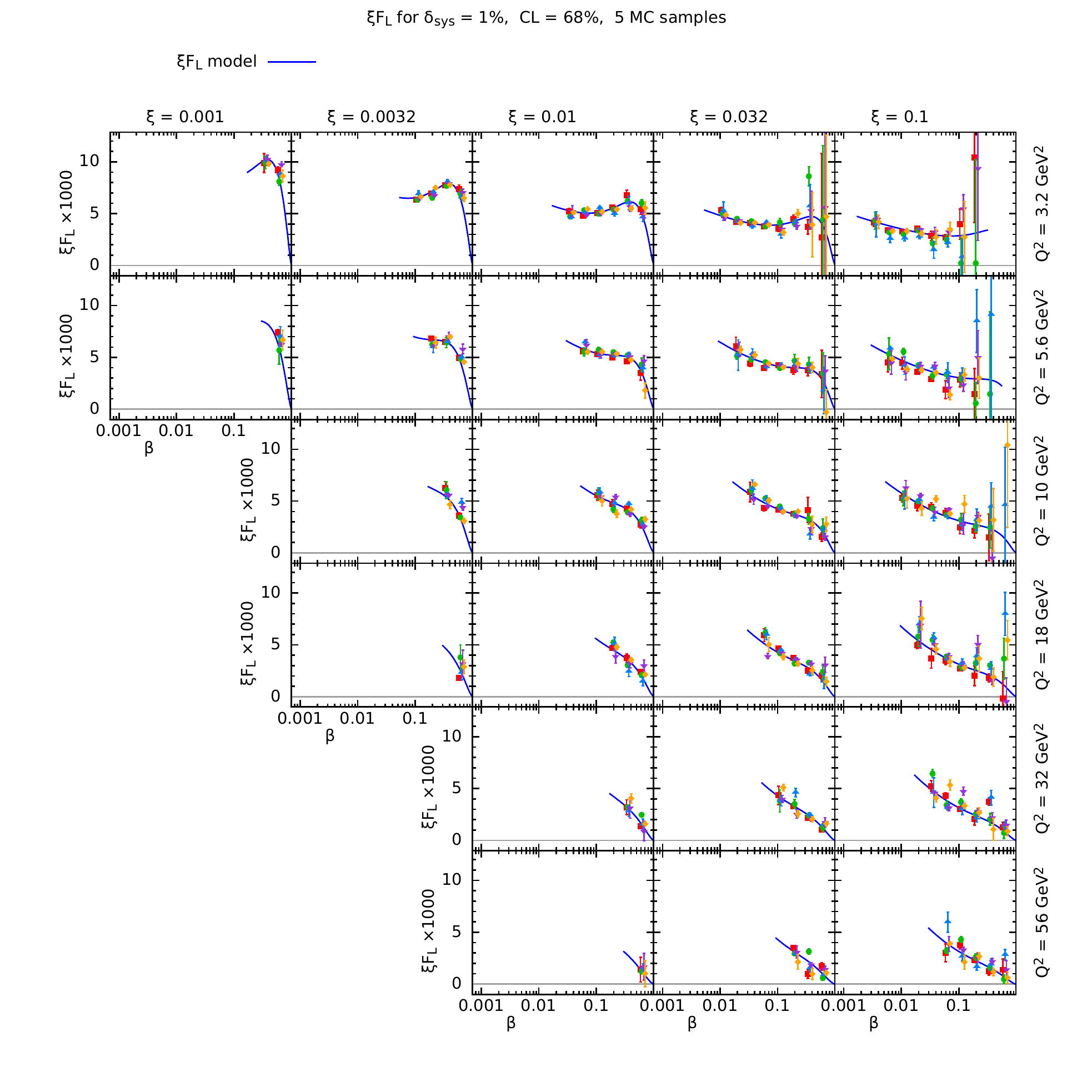}
}%}
\caption{Extracted \FLD for set \EsetAll, \(\esys\)=1\%, 68\% CL uncertainty bands, for 5 MC samples (markers with different colours, horizontally displaced 
from each other for clarity). The solid lines show the central values of the model used for the generation of the pseudodata from which \FLD was extracted.}
\label{fig:FLD5MC}
\end{figure}

For each choice of beam energy 
combinations and systematic precision, ten Monte Carlo simulations
are carried out in order to get a feel for the expected spread of
the results after propagation through the Rosenbluth fits.
Fig.~\ref{fig:compar_syst_ener} shows the extracted values of $\xi$\FLD as a function of $\beta$ for selected values of
$\xi$ and $Q^2$, with five randomly chosen 
examples of the Monte Carlo
replicas overlayed. 
In addition to the three different beam energy sets with 
$\esys\ = 1\%$ studied in 
Figs.~\ref{fig:FL_fit_AB} and~\ref{fig:FL_fit_AD}, results are
also shown for the \EsetAll set with $\esys\ = 2\%$.
The results for the full set of $(\beta, Q^2, \xi)$ bins 
with \EsetAll and $\esys\ = 1\%$ are
shown in Fig.~\ref{fig:FLD5MC}. 

There are not large differences between the 
results with the \EsetAll and \EsetBig beam energy sets: 
\EsetBig leads to a small 
reduction in the access to small $\beta$ values and some
increase of the uncertainties as measured by the spread between
different Monte Carlo sets. The differences are more 
pronounced in the comparison between \EsetAll and \EsetMin, due to the smaller number of bins in $(\xi,\beta,Q^2)$ accessible with \EsetMin.
In addition to larger uncertainties, the kinematic
range with \EsetMin is restricted to relatively larger $\beta$.
Therefore, although a larger number of energy combinations certainly yields better results, an extraction of \FLD is 
feasible for the EIC-favoured set of energy combinations.

As expected, the influence of a factor 2 in \(\esys\) 
translates into a factor around 2 in the uncertainties in \FLD.
This reflects the fact that with the assumed 
luminosity of 10 fb$^{-1}$ per collision energy,
systematic
uncertainties are the limiting factor throughout the kinematically accessible region.

    % --- WS:  FL in selected bins
    
\subsection{Estimated precision on \texorpdfstring{\FLD}{FLD}}
\label{subsec:MC}

Figs.~\ref{fig:compar_syst_ener} and \ref{fig:FLD5MC} 
illustrate the spread of
results for \FLD extracted from 
68\% CL uncertainties on fits to 
five different Monte Carlo 
pseudodata samples generated using the same central values
and uncertainties. 
The resulting distributions at each 
$(Q^2, \beta, \xi)$ point are complicated, since in addition  to 
the uncertainties propagated through the fits from 
the pseudodata, they also reflect the available $\YL$ range 
and number of data points available in the fits. 
In order to estimate the precision with which \FLD can be 
extracted at each $(Q^2, \beta, \xi)$ point, we
investigate the spread between the results obtained with the
different Monte Carlo samples, quantified 
using a direct arithmetic averaging procedure, 
neglecting the uncertainties obtained from the fits. The
mean and variance are therefore taken to be
\begin{equation}
\bar v = S_1/N, \quad (\Delta v)^2 = \frac {S_2 - S_1^2/N}{N-1}\ ,
\end{equation}
with $S_n = \sum_{i=1}^N v_i^n$ and $v_i$ the value of the extracted $\xi$\FLD in Monte Carlo sample $i$. 
The uncertainty is then taken to be $\Delta v$.

\begin{figure}[htb]
\includegraphics[width=0.9\columnwidth, clip, trim=0 0 0 40]{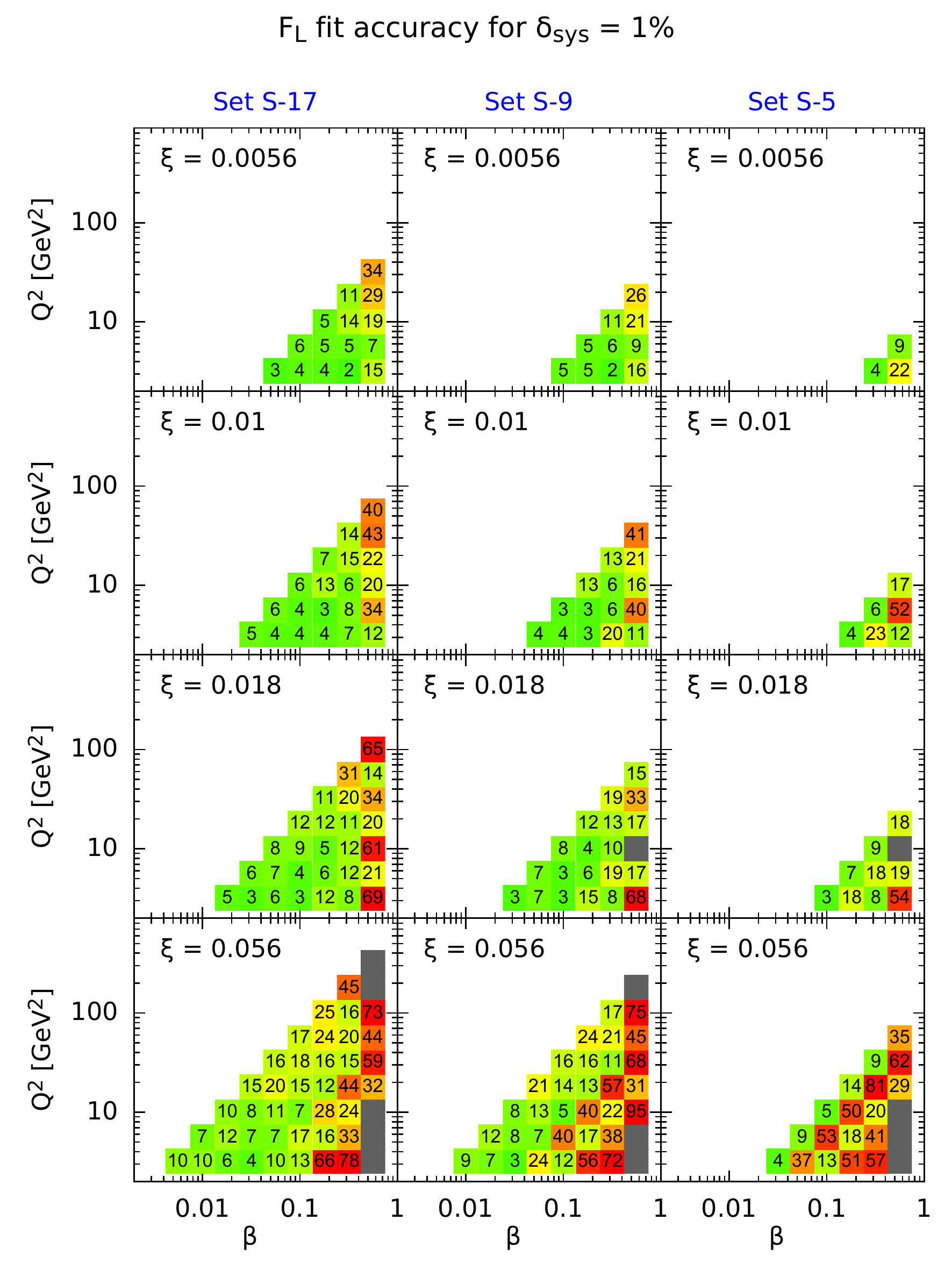}
\caption{Estimated uncertainties $\Delta v$ on \FLD 
extractions for beam energy sets \EsetAll, \EsetBig, \EsetMin, averaged over 10 MC samples for \esys = 1\%. The numbers in each box give the relative accuracy of the \FLD determination (in percent). Bins with errors exceeding 100\% are indicated in grey.}
\label{fig:uncer_arith_ave}
\end{figure}

Fig.~\ref{fig:uncer_arith_ave} shows the 
uncertainties $\Delta v$ in ($\beta, Q^2, \xi$) bins
for the three sets of beam energies, obtained by averaging over 10 separate 
Monte Carlo simulations. 
Even with 10 MC samples, there are some strong 
point-to-point fluctuations in the estimated uncertainties. 
However, the general trends become clear, in particular the regions
in which reliable measurements can be made.
The best measured region for each $\xi$ value is at the lowest 
$\beta$ and $Q^2$, where
the $\YL$ range is widest and statistical errors are negligible. 
The
uncertainties in a given bin do not depend strongly on the 
energy sets, but the accessible kinematic range 
in which measurements can be 
made decreases with decreasing numbers of energy combinations in the set.

%%%%%%%%%%%%%%%%%%%%%%%%%%%%%%%%%%%%%%
\subsection{Results for the ratio of the longitudinal to transverse structure functions}
\label{subsec:R}

The `photoabsorption' ratio in diffraction, 
$R^{\rm D(3)} = \FLD / \FTD$, where 
$\FTD = \FtwoD - \FLD$, is the
ratio of the cross section for longitudinally polarised photons to that for transversely polarised photons at the same $(\beta, Q^2, \xi)$ values. 
$R^{\rm D(3)}$ has a clear and intuitive physical meaning and can be
compared with similar quantities extracted from decay angular 
distributions in exclusive processes such as vector meson production.
It can be extracted from a fit to 
the reduced cross section pseudodata as a function of $\YL$ of the form
\begin{equation}
\sred^{\DD(3)} =  \left[ 1+\left(1 - \YL\right)\,R^{\rm D(3)} \right]
\FTD
\;,
\end{equation}
with $R^{\rm D(3)}$ and $\FTD$ being the free fit parameters. 
This alternative fit has different sensitivities to 
the uncertainties in the measurements from those in the $\FLD$
extraction.

        \begin{figure}[htb]
\centerline{%\fbox{
% height=0.8\columnwidth,
\includegraphics[width=0.9\columnwidth, clip,trim=0 29 0 58]{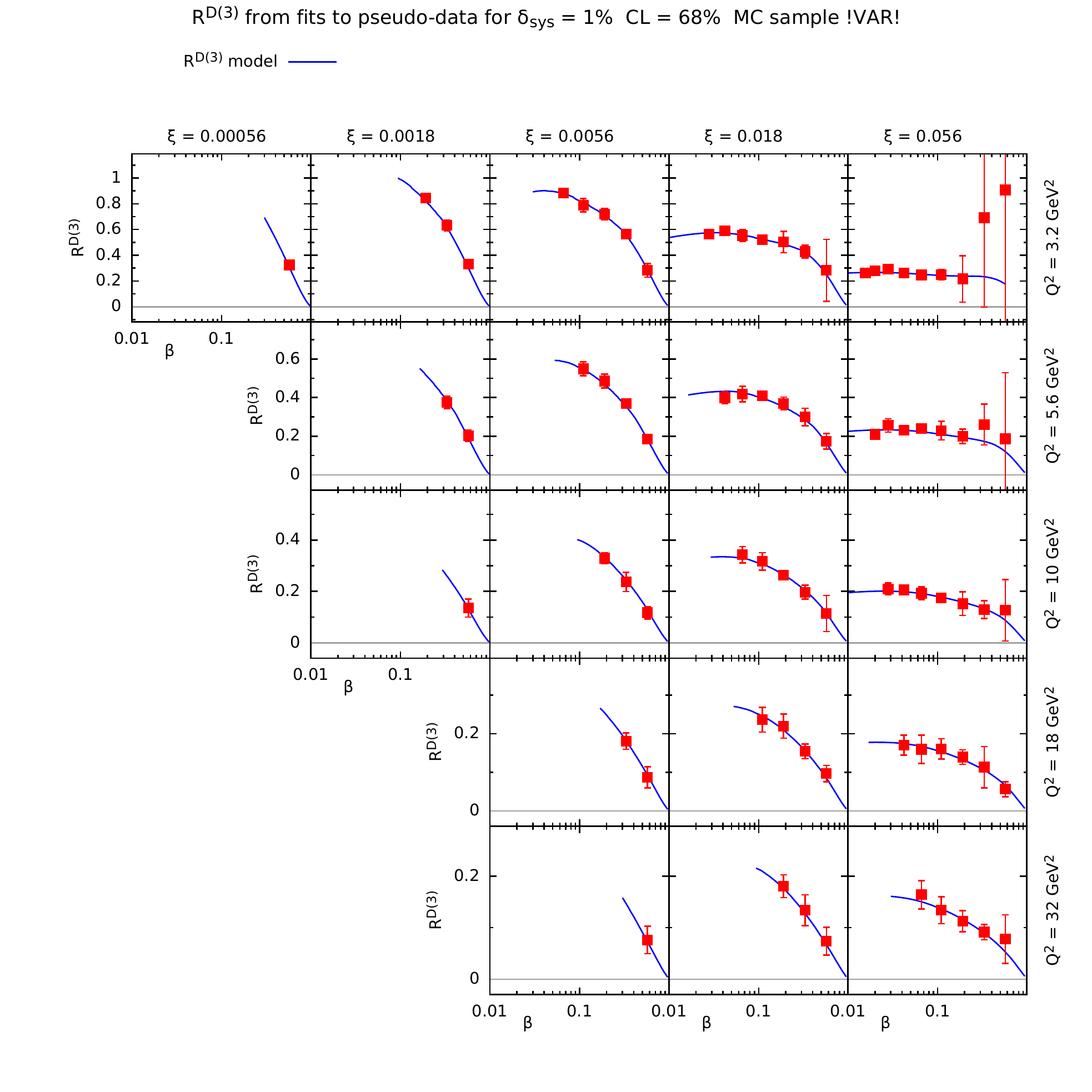}
}%}
\caption{Extracted $R^{\rm D(3)}$ data for beam energy 
set \EsetAll averaged over 10 MC samples with \esys = 1\%. 
The solid lines show the central values of the model used 
for the generation of the pseudodata. 
}
\label{fig:R}
\end{figure}
   
Fig.~\ref{fig:R} shows the results for the values of 
$R^{\rm D(3)}$ obtained from the averaging method 
described in Sec.~\ref{subsec:MC}
using 10 Monte Carlo samples and the largest set of beam energies,
\EsetAll and \esys = 1\%. A precise determination of
$R^{\rm D(3)}$ over a large kinematic range 
will be possible at the EIC.

\section{Conclusions and Outlook}
\label{sec:conclu}

In this paper we have investigated the potential of the Electron Ion Collider for the measurement of the longitudinal structure function in diffraction. This is a challenging measurement that requires data with high 
statistics at
several different centre-of-mass energies, ideally well beyond those
available in the pioneering measurement by the H1 collaboration at HERA.
We have considered 
EIC scenarios with 17, 9 and 5  different values of $\sqrt{s}$, the latter being the commonly assumed EIC scenario. 
Pseudodata for the reduced diffractive cross section 
were generated using a model based on 
collinear factorization with DGLAP evolution, assuming an integrated luminosity of  $10 \,\rm fb^{-1}$ for each centre-of-mass energy.
The uncorrelated systematic errors 
are assumed to be 
either $1 \%$ or $2 \%$, which is challenging compared with previous
measurements, but consistent with the expected high level of performance
of the EIC detectors. The longitudinal structure function was extracted using the standard Rosenbluth method of a linear fit to the 
reduced cross section as a function of the $\YL$ variable,
extracting $\FLD$ from the slope and $F_2^{\DD(3)}$
from the intercept.   
Fits were only performed in $(\beta,\xi,Q^2)$ bins where
at least 
four centre-of-mass energies yielded data points at accessible 
$y$ values
(in contrast to three in the H1 case). 
The scenarios with 17 and 9 centre-of-mass energies do not differ much in terms of the kinematic range in which $\FLD$ can be extracted, 
whereas the scenario with only 5 centre-of-mass energies 
results in a restricted kinematic range, particularly at 
small values of $\beta$ and $\xi$.  
The precision on the extracted value of $\FLD$ is strongly correlated with the available range in $\YL$ in a given bin $(\xi,\beta,Q^2)$ and 
is not diminished substantially when going from 17 to 9 centre-of-mass energies. A larger difference is observed 
for the set with 5 centre-of-mass energies,
mainly due to the smaller available range in $y$. 
Nonetheless, for bins in which measurements are possible, 
the precision of the extracted structure function \FLD 
is comparable in all scenarios. Given the very high target 
luminosity at the 
EIC, the exact choices of running time at each beam energy will not 
be a strongly limiting factor; the precision is likely to depend
much more strongly on the size of the systematic uncertainties that are
not correlated between data points.

We have also performed a separate extraction of the ratio 
$R^{\rm D(3)} = \FLD/(\FtwoD - \FLD)$ of the longitudinal to the transverse diffractive structure functions. A precise extraction of this quantity 
is expected to be possible at the EIC.

As an outlook, we note that the currently foreseen forward instrumentation at the EIC allows a precise determination of $-t$ in a wide range, see Sec.~\ref{subsec:ptag}. It will be very interesting to explore to what extent an extraction of $F_\mathrm{L}^{\DD(4)}(\beta,\xi,Q^2,t)$ in a similar way to that illustrated previously for \FLD, Eq.~(\ref{eq:sred4}), will be possible. This is a completely new study that we leave for the future.

% ++++++++++++++++++++++++++++++++++++++++++++++++++++++++++++++++++

\section*{Acknowledgements}
We thank Alex Jentsch, Hannes Jung, Kolja Kauder, and Mark Strikman for useful discussions.
NA acknowledges financial support by Xun\-ta de Galicia (Centro singular de investigaci\'on de Galicia accreditation 2019-2022); the "Mar\'{\i}a de Maeztu" Units of Excellence program MDM2016-0692 and the Spanish Research State Agency under project FPA2017-83814-P; European Union ERDF; the European Research Council under project
ERC-2018-ADG-835105 YoctoLHC; MSCA RISE 823947 "Heavy ion collisions: collectivity and precision in saturation physics"
(HIEIC); and European Union's Horizon 2020 research and innovation programme under
grant agreement No. 824093.
AMS is  supported by the U.S. Department of Energy grant No. DE-SC-0002145 and in part by National Science Centre in Poland, grant 2019/33/B/ST2/02588.

%\bibliography{mybib}

\begin{thebibliography}{10}

\bibitem{Adloff:1997sc}
C.~Adloff et~al.
\newblock {Inclusive measurement of diffractive deep inelastic ep scattering}.
\newblock {\em Z. Phys.}, C76:613--629, 1997.

\bibitem{Breitweg:1997aa}
J.~Breitweg et~al.
\newblock {Measurement of the diffractive structure function $F_2^{D(4)}$ at
  HERA}.
\newblock {\em Eur. Phys. J.}, C1:81--96, 1998.

\bibitem{Newman:2013ada}
P.~Newman and M.~Wing.
\newblock {The Hadronic Final State at HERA}.
\newblock {\em Rev. Mod. Phys.}, 86(3):1037, 2014.

\bibitem{Aktas:2006hx}
A.~Aktas et~al.
\newblock {Diffractive deep-inelastic scattering with a leading proton at
  HERA}.
\newblock {\em Eur.Phys.J.}, C48:749--766, 2006.

\bibitem{Aaron:2010aa}
F.~D. Aaron et~al.
\newblock {Measurement of the cross section for diffractive deep-inelastic
  scattering with a leading proton at HERA}.
\newblock {\em Eur. Phys. J. C}, 71:1578, 2011.

\bibitem{ZEUS:2004luu}
S.~Chekanov et~al.
\newblock {Dissociation of virtual photons in events with a leading proton at
  HERA}.
\newblock {\em Eur. Phys. J. C}, 38:43--67, 2004.

\bibitem{ZEUS:2008xhs}
S.~Chekanov et~al.
\newblock {Deep inelastic scattering with leading protons or large rapidity
  gaps at HERA}.
\newblock {\em Nucl. Phys. B}, 816:1--61, 2009.

\bibitem{H1:2006zyl}
A.~Aktas et~al.
\newblock {Measurement and QCD analysis of the diffractive deep-inelastic
  scattering cross-section at HERA}.
\newblock {\em Eur. Phys. J. C}, 48:715--748, 2006.

\bibitem{H1:2012pbl}
F.~D. Aaron et~al.
\newblock {Inclusive Measurement of Diffractive Deep-Inelastic Scattering at
  HERA}.
\newblock {\em Eur. Phys. J. C}, 72:2074, 2012.

\bibitem{Chekanov:2009aa}
S.~Chekanov et~al.
\newblock {A QCD analysis of ZEUS diffractive data}.
\newblock {\em Nucl.Phys.}, B831:1--25, 2010.

\bibitem{Kaidalov:1979jz}
A.~B. Kaidalov.
\newblock {Diffractive Production Mechanisms}.
\newblock {\em Phys. Rept.}, 50:157--226, 1979.

\bibitem{Kaidalov:2003vg}
A.~B. Kaidalov, V.~A. Khoze, A.~D. Martin, and M.~G. Ryskin.
\newblock {Diffraction of protons and nuclei at high-energies}.
\newblock {\em Acta Phys. Polon. B}, 34:3163--3190, 2003.

\bibitem{Bartels:2000ze}
J.~Bartels and H.~Kowalski.
\newblock {Diffraction at HERA and the confinement problem}.
\newblock {\em Eur. Phys. J. C}, 19:693--708, 2001.

\bibitem{Kovchegov:2012mbw}
Y.~V. Kovchegov and E.~Levin.
\newblock {\em {Quantum chromodynamics at high energy}}, volume~33.
\newblock Cambridge University Press, 8 2012.

\bibitem{Gribov:1968jf}
V.~N. Gribov.
\newblock {Glauber corrections and the interaction between high-energy hadrons
  and nuclei}.
\newblock {\em Sov. Phys. JETP}, 29:483--487, 1969.
\newblock [Zh. Eksp. Teor. Fiz.56,892(1969)].

\bibitem{Collins:1997sr}
J.~C. Collins.
\newblock {Proof of factorization for diffractive hard scattering}.
\newblock {\em Phys. Rev.}, D57:3051--3056, 1998.
\newblock [Erratum: Phys. Rev.D61,019902(2000)].

\bibitem{Berera:1995fj}
A.~Berera and D.~E. Soper.
\newblock {Behavior of diffractive parton distribution functions}.
\newblock {\em Phys. Rev.}, D53:6162--6179, 1996.

\bibitem{Trentadue:1993ka}
L.~Trentadue and G.~Veneziano.
\newblock {Fracture functions: An Improved description of inclusive hard
  processes in QCD}.
\newblock {\em Phys. Lett.}, B323:201--211, 1994.

\bibitem{Klasen:2008ah}
M.~Klasen and G.~Kramer.
\newblock {Review of factorization breaking in diffractive photoproduction of
  dijets}.
\newblock {\em Mod. Phys. Lett. A}, 23:1885--1907, 2008.

\bibitem{LHCForwardPhysicsWorkingGroup:2016ote}
K.~Akiba et~al.
\newblock {LHC Forward Physics}.
\newblock {\em J. Phys. G}, 43:110201, 2016.

\bibitem{Motyka:2012ty}
L.~Motyka, M.~Sadzikowski, and W.~Slominski.
\newblock {Evidence of strong higher twist effects in diffractive DIS at HERA
  at moderate $Q^2$}.
\newblock {\em Phys. Rev.}, D86:111501, 2012.

\bibitem{Aaron:2012zz}
F.~D. Aaron et~al.
\newblock {Measurement of the Diffractive Longitudinal Structure Function
  $F_L^D$ at HERA}.
\newblock {\em Eur. Phys. J. C}, 71:1836, 2011.

\bibitem{Accardi:2012qut}
A.~Accardi et~al.
\newblock {Electron Ion Collider: The Next QCD Frontier}.
\newblock {\em Eur. Phys. J.}, A52(9):268, 2016.

\bibitem{AbdulKhalek:2021gbh}
R.~Abdul~Khalek et~al.
\newblock {Science Requirements and Detector Concepts for the Electron-Ion
  Collider: EIC Yellow Report}.
\newblock 3 2021.

\bibitem{AbelleiraFernandez:2012cc}
J.~L. Abelleira~Fernandez et~al.
\newblock {A Large Hadron Electron Collider at CERN: Report on the Physics and
  Design Concepts for Machine and Detector}.
\newblock {\em J. Phys.}, G39:075001, 2012.

\bibitem{LHeC:2020van}
P.~Agostini et~al.
\newblock {The Large Hadron-Electron Collider at the HL-LHC}.
\newblock 7 2020.

\bibitem{FCC:2018byv}
A.~Abada et~al.
\newblock {FCC Physics Opportunities}: {Future Circular Collider Conceptual
  Design Report Volume 1}.
\newblock {\em Eur. Phys. J. C}, 79(6):474, 2019.

\bibitem{FCC:2018vvp}
A.~Abada et~al.
\newblock {FCC-hh: The Hadron Collider}: {Future Circular Collider Conceptual
  Design Report Volume 3}.
\newblock {\em Eur. Phys. J. ST}, 228(4):755--1107, 2019.

\bibitem{Armesto:2019gxy}
N.~Armesto, P.~R. Newman, W.~S\l{}omi\'nski, and A.~M. Sta\'sto.
\newblock {Inclusive diffraction in future electron-proton and electron-ion
  colliders}.
\newblock {\em Phys. Rev. D}, 100(7):074022, 2019.

\bibitem{Slominski:2021zit}
W.~Slominski, N.~Armesto, P.~R. Newman, and A.~Stasto.
\newblock {Opportunities for inclusive diffraction at EIC}.
\newblock 7 2021.

\bibitem{Chekanov:2005vv}
S.~Chekanov et~al.
\newblock {Study of deep inelastic inclusive and diffractive scattering with
  the ZEUS forward plug calorimeter}.
\newblock {\em Nucl. Phys.}, B713(1-3):3--80, 2005.

\bibitem{Aktas:2006hy}
A.~Aktas et~al.
\newblock {Measurement and QCD analysis of the diffractive deep-inelastic
  scattering cross-section at HERA}.
\newblock {\em Eur.Phys.J.}, C48:715--748, 2006.

\bibitem{Chekanov:2008fh}
S.~Chekanov et~al.
\newblock {Deep inelastic scattering with leading protons or large rapidity
  gaps at HERA}.
\newblock {\em Nucl.Phys.}, B816:1--61, 2009.

\bibitem{Aaron:2012ad}
F.D. Aaron et~al.
\newblock {Inclusive Measurement of Diffractive Deep-Inelastic Scattering at
  HERA}.
\newblock {\em Eur.Phys.J.}, C72:2074, 2012.

\bibitem{Gribov:1972rt}
V.~N. Gribov and L.~N. Lipatov.
\newblock {$e^+ \, e^-$ pair annihilation and deep inelastic e p scattering in
  perturbation theory}.
\newblock {\em Sov. J. Nucl. Phys.}, 15:675--684, 1972.
\newblock [Yad. Fiz.15,1218(1972)].

\bibitem{Gribov:1972ri}
V.~N. Gribov and L.~N. Lipatov.
\newblock {Deep inelastic e p scattering in perturbation theory}.
\newblock {\em Sov. J. Nucl. Phys.}, 15:438--450, 1972.
\newblock [Yad. Fiz.15,781(1972)].

\bibitem{Altarelli:1977zs}
G.~Altarelli and G.~Parisi.
\newblock {Asymptotic Freedom in Parton Language}.
\newblock {\em Nucl. Phys.}, B126:298--318, 1977.

\bibitem{Dokshitzer:1977sg}
Y.~L. Dokshitzer.
\newblock {Calculation of the Structure Functions for Deep Inelastic Scattering
  and $e^+ \, e^-$ Annihilation by Perturbation Theory in Quantum
  Chromodynamics.}
\newblock {\em Sov. Phys. JETP}, 46:641--653, 1977.
\newblock [Zh. Eksp. Teor. Fiz.73,1216(1977)].

\bibitem{ZEUS:1996bgb}
M.~Derrick et~al.
\newblock {Study of elastic $\rho^0$ photoproduction at HERA using the ZEUS
  leading proton spectrometer}.
\newblock {\em Z. Phys. C}, 73:253--268, 1997.

\bibitem{VanEsch:1999pi}
P.~Van~Esch et~al.
\newblock {The H1 forward proton spectrometer at HERA}.
\newblock {\em Nucl. Instrum. Meth. A}, 446:409--425, 2000.

\bibitem{Astvatsatourov:2014dna}
A.~Astvatsatourov, K.~Cerny, J.~Delvax, L.~Favart, T.~Hreus, X.~Janssen,
  R.~Roosen, T.~Sykora, and P.~Van~Mechelen.
\newblock {The H1 very forward proton spectrometer at HERA}.
\newblock {\em Nucl. Instrum. Meth. A}, 736:46--65, 2014.

\bibitem{AbdelKhalek:2016tiv}
S.~Abdel~Khalek et~al.
\newblock {The ALFA Roman Pot Detectors of ATLAS}.
\newblock {\em JINST}, 11(11):P11013, 2016.

\bibitem{Adamczyk:2017378}
L~Adamczyk, E~Banaś, A~Brandt, M~Bruschi, S~Grinstein, J~Lange, M~Rijssenbeek,
  P~Sicho, R~Staszewski, T~Sykora, M~Trzebiński, J~Chwastowski, and K~Korcyl.
\newblock {Technical Design Report for the ATLAS Forward Proton Detector}.
\newblock Technical report, May 2015.

\bibitem{TOTEM:2017asr}
G.~Antchev et~al.
\newblock {First measurement of elastic, inelastic and total cross-section at
  $\sqrt{s}=13$ TeV by TOTEM and overview of cross-section data at LHC
  energies}.
\newblock {\em Eur. Phys. J. C}, 79(2):103, 2019.

\bibitem{CMS:2018uvs}
Albert~M Sirunyan et~al.
\newblock {Observation of proton-tagged, central (semi)exclusive production of
  high-mass lepton pairs in pp collisions at 13 TeV with the CMS-TOTEM
  precision proton spectrometer}.
\newblock {\em JHEP}, 07:153, 2018.

\bibitem{ZEUS:2009uxs}
S.~Chekanov et~al.
\newblock {A QCD analysis of ZEUS diffractive data}.
\newblock {\em Nucl. Phys. B}, 831:1--25, 2010.

\bibitem{Collins:1986mp}
J.~C. Collins and W.-K. Tung.
\newblock {Calculating Heavy Quark Distributions}.
\newblock {\em Nucl. Phys.}, B278:934, 1986.

\bibitem{Thorne:2008xf}
R.~S. Thorne and W.~K. Tung.
\newblock {PQCD Formulations with Heavy Quark Masses and Global Analysis}.
\newblock 2008.

\bibitem{Owens:1984zj}
J.~F. Owens.
\newblock {$Q^2$ Dependent Parametrizations of Pion Parton Distribution
  Functions}.
\newblock {\em Phys. Rev.}, D30:943, 1984.

\bibitem{Gluck:1991ey}
M.~Gluck, E.~Reya, and A.~Vogt.
\newblock {Pionic parton distributions}.
\newblock {\em Z. Phys.}, C53:651--656, 1992.

\bibitem{Jung:1993gf}
H.~Jung.
\newblock {Hard diffractive scattering in high-energy e p collisions and the
  Monte Carlo generator RAPGAP}.
\newblock {\em Comput. Phys. Commun.}, 86:147--161, 1995.

\bibitem{H1:2008rkk}
F.~D. Aaron et~al.
\newblock {Measurement of the Proton Structure Function $F_L$(x, $Q^2$) at Low
  x}.
\newblock {\em Phys. Lett. B}, 665:139--146, 2008.

\bibitem{ZEUS:2009nwk}
S.~Chekanov et~al.
\newblock {Measurement of the Longitudinal Proton Structure Function at HERA}.
\newblock {\em Phys. Lett. B}, 682:8--22, 2009.

\bibitem{H1:2010fzx}
F.~D. Aaron et~al.
\newblock {Measurement of the Inclusive ${e^\pm}$p Scattering Cross Section at
  High Inelasticity y and of the Structure Function $F_L$}.
\newblock {\em Eur. Phys. J. C}, 71:1579, 2011.

\end{thebibliography}

\end{document}